\documentclass[sigconf]{acmart}
\AtBeginDocument{%
  }

\usepackage[normalem]{ulem}
\usepackage{booktabs} 
\usepackage{subcaption}

\usepackage[english]{babel}
\usepackage[utf8x]{inputenc}
\usepackage[T1]{fontenc}
\usepackage{comment} 

\usepackage{amsmath}
\usepackage{graphicx}
\usepackage[table]{xcolor}
\usepackage[colorinlistoftodos]{todonotes}
\usepackage{multirow}
\usepackage{array} 
\usepackage{mwe}
\usepackage{xcolor, soul}
\sethlcolor{red}
\usepackage{siunitx}

\begin{document}

\title{Multi-Objective Evolutionary Design of Molecules with Enhanced Nonlinear Optical Properties}



\author{Dominic Mashak}
\affiliation{%
 \institution{Southwestern University}
 \city{Georgetown}
 \state{Texas}
 \country{USA}}
\email{mashakd@southwestern.edu}

\author{Jacob Schrum}
\affiliation{%
 \institution{Southwestern University}
 \city{Georgetown}
 \state{Texas}
 \country{USA}}
\email{schrum2@southwestern.edu}

\author{S.A. Alexander}
\affiliation{%
 \institution{Southwestern University}
 \city{Georgetown}
 \state{Texas}
 \country{USA}}
\email{alexands@southwestern.edu}

\begin{abstract}
Nonlinear optical (NLO) materials are essential for many photonic, telecommunication, and laser technologies, yet discovering better NLO molecules is computationally challenging due to the vast chemical space and competing objectives. We compare evolutionary algorithms for molecular design, targeting four objectives: maximizing the ratio of first-to-second hyperpolarizability $(\beta/\gamma)$, optimizing HOMO-LUMO gap and linear polarizability to target ranges, and minimizing energy per atom. We encode molecules as 
SMILES strings and evaluate their properties using quantum-chemical calculations. 
We compare
NSGA-II,
MAP-Elites,
MOME, a single-objective $(\mu+\lambda)$ evolutionary algorithm, and simulated annealing. Quality diversity methods maintain archives across a measure space defined by atom and bond count, enabling the discovery of structurally diverse molecules. Our results demonstrate that NSGA-II consistently earns high scores in every objective, leading to high-quality molecules,
but MOME does a better job exploring a wide range of possibilities, resulting in higher global hypervolume and MOQD scores. However, each method has strengths and weaknesses, and produced many promising molecules.
\end{abstract}

\ccsdesc[500]{Applied computing~Chemistry}
\ccsdesc[500]{Computing methodologies~Molecular simulation}

\keywords{Multi-Objective optimization, Quality Diversity, Computational Chemistry}

\settopmatter{printacmref=false}
\renewcommand\footnotetextcopyrightpermission[1]{}
\pagestyle{plain}

\maketitle

\setlength{\tabcolsep}{3pt}

\section{Introduction}

Nonlinear optical (NLO) materials can modify the frequency, phase, and/or polarization of light~\cite{minasian:elsevier2005, piela:elsevier2020, saleh:photonics1991}. Because these properties enable optical communication, optical computing, optical data storage, and optical switching devices, many research groups are actively seeking molecules with specific NLO properties~\cite{dalton:spie2004, ros:springer2008, dastmalchi:wp2014}. Over the past few decades, the development of reliable black-box quantum-chemical programs has enabled theoretical chemists to evaluate the properties of NLO molecules solely from their structures.

Despite the commercial importance of NLO materials, the scientific literature contains only general guidelines as to which properties an ideal NLO material should have for a specific application. Most studies focus on
a particular property, such as molecular first hyperpolarizability ($\beta$)~\cite{kanis:cr1994, marder:science1991, singer:josab1989, philip:acr2010, hrivnak:molecules2022}. In this paper, we consider an electro-optic modulator~\cite{minasian:elsevier2005} as an example NLO device, and identify properties that a molecule must possess to function as an electro-optic modulator (Section~\ref{sec:electrooptic}). 
We seek molecules that optimize these properties using multiobjective optimization (MOO)~\cite{deb:tec2002}, quality diversity (QD) \cite{mouret:arxiv2015},
multiobjective QD (MOQD)~\cite{pierrot:gecco2022}, single-objective evolution, and simulated annealing. 
Previous work~\cite{mashak:gecco2024} demonstrates that evolutionary algorithms can be used to search for molecules with large hyperpolarizabilities, but we apply a wider variety of approaches to a harder problem:
finding molecules that simultaneously satisfy multiple, often conflicting, criteria. 



Our results show that MOQD via the MOME algorithm~\cite{pierrot:gecco2022} 
creates the best variety of molecules,
covering many diverse niches while also
maximizing hypervolume~\cite{zitzler:tec2003}, but the
multiobjective optimization algorithm NSGA-II~\cite{deb:tec2002}
uncovers higher scores in each individual objective, providing a different set
of trade-offs. Single-objective methods optimize our main objective
of first-to-second hyperpolarizability ratio at the expense of other
objectives, but the diversity that is fostered 
by the QD method MAP-Elites \cite{mouret:arxiv2015} allows it
to perform better on a wider range of objectives, even though it is not aware of them.
All of these approaches provide an interesting variety of potential NLO
materials for further study.


\section{Related Work}


Various search methods have been 
applied to molecular design.
Simulated annealing and evolutionary algorithms have been used to optimize molecular hyperpolarizabilities with semi-empirical quantum chemistry~\cite{mashak:match2024, mashak:gecco2024}, 
representing molecules with SMILES strings~\cite{weininger:jcics1988}, which we also use in our experiments (Section \ref{sec:smiles}).

Though effective, such optimization techniques generally ignore other aspects of design that impact solution usefulness. In contrast, quality diversity (QD) approaches~\cite{mouret:arxiv2015,cully:nature15}
seek diverse collections of artifacts while also maximizing fitness. Graph-based elite patch illumination (GB-EPI~\cite{verhellen:rsc2020}) is a QD approach 
that was applied to small-molecule drug rediscovery benchmarks. GB-EPI evolves graph-based representations instead of SMILES strings, while our experiments combine SMILES with QD.

A separate limitation of standard optimization is its focus on a single objective.
Multiobjective Pareto-based optimization 
addresses this by exploring trade-offs between competing objectives rather than
seek diversity in design space, and
has also been applied to evolving
molecules for small-molecule drug design~\cite{homberg:cec2024, verhellen:cs2022}.





We apply QD and multiobjective (MO) techniques, along with others, to the problem of discovering molecules for the design of an effective electro-optic modulator.

\section{Defining an Effective Electro-Optic Modulator}
\label{sec:electrooptic}

When intense laser light strikes an electro-optic (EO) modulator, the molecular response to the electric field~\cite{piela:elsevier2020} determines output properties via polarizability tensors $\alpha$, $\beta$, and $\gamma$. An effective EO modulator requires balancing four properties. First, EO modulators exploit the Pockels effect~\cite{saleh:photonics1991}, a second-order NLO process proportional to hyperpolarizability ($\beta$); high $\beta$ enables stronger modulation and smaller devices, while $\gamma$ should be small enough to avoid self-phase modulation and optical Kerr effects~\cite{saleh:photonics1991} yet sufficient for intense-field performance~\cite{saleh:photonics1991}. Second, linear polarizability ($\alpha$) must be balanced: high values provide strong charge transfer (beneficial for $\beta$ and $\gamma$), but excessive $\alpha$ causes absorption, dispersion, or aggregation. Third, the HOMO-LUMO gap ($\Delta E$) controls electron mobility and optical transparency; small gaps risk visible-light absorption and thermal damage, while large gaps ensure transparency but weaken NLO activity. Fourth, molecules must be thermodynamically stable. These properties are calculable via quantum-chemical programs, which we combine with evolutionary algorithms to search for optimal EO modulators.

\section{Methods}

SMILES strings representing molecules are evolved using MOO, QD, MOQD, and single-objective evolution, and compared with simulated annealing. 
The properties of candidate molecules are calculated using the PySCF library \cite{sun:jcphys2020}.



\subsection{SMILES String Encoding}
\label{sec:smiles}

SMILES (Simplified Molecular Input Line Entry System~\cite{weininger:jcics1988}) encodes molecular structures as ASCII strings. We restrict molecules to C, N, O, and H atoms, focusing on organic NLO candidates, following prior work~\cite{mashak:gecco2024}.
Atoms are represented by atomic symbols; organic atoms (C, O) omit brackets and use implicit hydrogens based on valence. 
Bonds are limited to single (implicit \texttt{-}) and double (\texttt{=}) bonds. 
Branches are represented by parentheses at the attachment point (e.g., C-C(-O)-O), and rings by matching digit labels (e.g., C1-C-C-C-C1).
Canonical SMILES from RDKit~\cite{landrum:rdkit2010} ensure unique representations for comparison and deduplication.


This encoding enables our integration with quantum chemistry packages 
for evaluation, as SMILES strings can be directly converted to molecular geometries for calculations of hyperpolarizabilities and other properties.

\subsection{Calculating Chemical Properties}
\label{sec:calcchem}

Candidate molecules are evaluated by converting canonical SMILES to molecular geometries using RDKit~\cite{landrum:rdkit2010}, then computing quantum-chemical properties with PySCF~\cite{sun:jcphys2020}, a Python library for \emph{ab initio} electronic structure calculations via Hartree-Fock (HF).

We use HF/3-21G based on systematic benchmarking~\cite{mashak:arxiv2025} showing perfect pairwise ranking agreement (100\% consistency ordering molecules by $\beta$) against experimental hyperpolarizabilities~\cite{kanis:cr1994}. While absolute errors exist, preserving relative ordering is critical for elitist/tournament selection in evolutionary algorithms.


Computed values yield the four objectives in Section~\ref{sec:objectives}. Molecules with invalid geometries or SCF failures receive the worst fitness values and are excluded per bounds in Section~\ref{sec:medianglobalhv}.

\subsection{Multiobjective Optimization}

Molecular design inherently involves competing objectives. 
Rather than optimize objectives independently or combine them, 
multiobjective optimization \cite{zitzler:ppsn1998} simultaneously optimizes multiple objectives to produce a set of trade-off solutions.

A solution $\vec{x}_1$ Pareto dominates $\vec{x}_2$ ($\vec{x}_1 \prec \vec{x}_2$) if $\vec{x}_1$ is no worse than $\vec{x}_2$ in every objective and strictly better in at least one. 
\begin{equation}
\vec{x}_1 \prec \vec{x}_2 \quad \Leftrightarrow \quad \forall i:\ f_i(\vec{x}_1) \leq f_i(\vec{x}_2) \;\; \land \;\; \exists j:\ f_j(\vec{x}_1) < f_j(\vec{x}_2)
\end{equation}


\noindent
This definition assumes minimization, but maximization is achieved by flipping inequalities.
The Pareto front is the set of all non-dominated solutions, representing the optimal trade-off surface between objectives.





\subsection{Quality Diversity}

Quality diversity (QD) algorithms seek not just high-performing solutions, but a diverse collection of high-performing solutions exhibiting different properties \cite{mouret:arxiv2015}. Rather than converge to a single optimum or Pareto front, QD methods generally maintain an archive of solutions that excel across various niches of design space.


Formally \cite{fontaine:nips21}, QD methods depend on a fitness function
and $k$ measure functions $m_{i} : \mathcal{S} \rightarrow \mathbb{R}$.
$\mathcal{S}$ is the space of candidate solutions (SMILES strings).
Each measure function calculates a solution property relevant to its design, but distinct from fitness.
A $k$-dimensional archive is discretized into bins with a chosen level of granularity, determining how novel a solution has to be in order to compete within a different evolutionary niche.

The measure functions determine where a solution fits in the archive.
Each bin holds one solution; higher fitness solutions replace worse ones, but solutions that discover empty bins are accepted regardless of fitness. QD algorithms maximize both coverage of the archive and overall quality of solutions. 
However, standard QD assumes only one fitness function determines solution quality.

\subsection{Multiobjective Quality Diversity}

MOQD algorithms \cite{pierrot:gecco2022} combine the complementary strengths of MOO and QD. While MOO methods optimize toward a Pareto front without regard to diversity in measure space, and QD algorithms maintain diversity but optimize only a single objective, MOQD pursues both goals: discovering diverse solutions that represent optimal objective trade-offs within each niche.





MOQD maintains an archive across the measure space while preserving multiple Pareto-optimal solutions in each niche.
The main modification to standard QD is allowing each archive bin to store a Pareto front \cite{pierrot:gecco2022} consisting of mutually non-dominated solutions that share measure characteristics, but represent different objective trade-offs.





\section{Experiment}

Our code is available on GitHub at \\ https://github.com/DominicMashak/Molecular-Evolution. These experiments focus on comparing different approaches to
evolving molecules for their NLO properties.




\subsection{Algorithms}
\label{sec:algo}


We compare five algorithms encompassing single-objective, MO, QD, and MOQD approaches.

Non-dominated Sorting Genetic Algorithm II (NSGA-II)~\cite{deb:tec2002} uses Pareto dominance ranking and crowding distance for selection. Solutions are sorted into non-dominated fronts, with binary tournament selection favoring better fronts. Crowding distance promotes spread along the Pareto front by preferring solutions in less populated objective space regions.

Multi-dimensional Archive of Phenotypic Elites (MAP-Elites)~\cite{mouret:arxiv2015} maintains an archive of elite solutions across a discretized measure space. 
Each bin stores the fittest solution for its niche, maintaining diversity via spatial partitioning.
During evolution, candidates are generated via mutation of randomly selected archive members, 
and assigned to bins based on their measures.
If a bin is empty, the candidate fills it. If occupied, the candidate replaces the current occupant only if it has superior fitness.
This process continues and fills the archive with diverse, high-quality solutions.

Multiobjective MAP-Elites (MOME)~\cite{pierrot:gecco2022} extends MAP-Elites by storing local Pareto fronts within each archive bin rather than single elites. Each bin maintains a mutually non-dominated Pareto front of objective trade-offs. 
Solutions are generated as with MAP-Elites, and assigned to bins based on their measures. They remain in their assigned bin if it is empty, or if they are non-dominated with respect to the previous occupants. Old occupants are only discarded if they are dominated by new occupants.

Simple $(\mu+\lambda)$ elitist selection~\cite{beyer:nc2002} is a single-objective method where $\mu$ parents generate $\lambda$ offspring per generation, and the best $\mu$ individuals from the combined parent-offspring pool are retained for the next generation. This elitist strategy is a simple baseline.

Simulated Annealing (SA)~\cite{dowsland:johnwiley1993, laarhoven:springer1987} is a non-evolutionary method that maintains a single solution, accepting improvements unconditionally and worse solutions with probability $\exp(\Delta E / T)$ via the Metropolis criterion~\cite{metropolis:jcphys1953}, where $\Delta E$ is the fitness difference and $T$ is the \emph{temperature}. Large $T$ values promote exploration while small values favor exploitation.


\subsection{Objectives}
\label{sec:objectives}

MO approaches target four objectives:

\begin{itemize}
\item {First-to-Second Hyperpolarizability Ratio ($\beta/\gamma$)}: We maximize this ratio to favor strong second-order responses relative to third-order effects, promoting efficient frequency conversion without competing third-order processes. $\beta$ is computed via second-order finite differences on dipole moments at 27 field configurations ($\pm h=0.001$ a.u.), yielding all 27 $\beta_{ijk}$ tensor components~\cite{kurtz:jcc1990, bishop:jws1998, mashak:arxiv2025}. Similarly, $\gamma$ uses fourth-order finite differences on total energy at field strengths $0, \pm h, \pm 2h$ along each axis; mean $\gamma$ averages the three diagonal components $\gamma_{iiii}$.

\item {Linear Polarizability Range Deviation ($f_{\alpha}$)}: Large $\alpha$ indicates high electronic mobility, but excessive values cause optical losses and reduced photostability~\cite{piela:elsevier2020}. We compute $\alpha$ via central finite differences on dipole moments with fields $\pm h$ along each axis: $\alpha_{ii} \approx [\mu_i(+h) - \mu_i(-h)] / (2h)$, then average tensor components. Targeting 100--500 a.u. (common in NLO materials with strong $\beta$), we minimize:
\begin{equation}
    f_\alpha = \max(0, 100 - \alpha) + \max(0, \alpha - 500)
\end{equation}

\item {HOMO-LUMO Gap Range Deviation ($f_{\Delta E}$)}: Gaps near 2 eV provide strong charge-transfer and high $\beta$ but risk red/near-IR absorption; gaps above 4 eV ensure visible transparency but reduce $\beta$. Targeting 2--4 eV for visible/near-IR applications, we minimize:
\begin{equation}
f_{\Delta E} = \max(0, 2 - \Delta E) + \max(0, \Delta E - 4)
\end{equation}

\item {Energy per Atom ($E_{total}/N_{atoms}$)}: Total energy from PySCF's geometry-optimized HF calculation, divided by atom count, proxies thermodynamic stability. Minimizing this encourages compact, low-strain geometries resisting decomposition, practical for synthesizable NLO materials. Positive values indicate SCF failures or strained geometries, so we reject those molecules (see Section \ref{sec:medianglobalhv} and \ref{sec:discussion}).
\end{itemize}

NSGA-II and MOME optimize all objectives; $(\mu+\lambda)$, MAP-Elites, and simulated annealing optimize only $\beta/\gamma$. This isolates the primary NLO metric to test whether high second-order responses correlate with desirable secondary objectives or require MOO. We avoid weighted sums due to scaling and hyperparameter sensitivity.

\subsection{Diversity Measures and Archives}


The archives used by MAP-Elites and MOME to characterize molecular diversity are based on two measures:
\begin{itemize}
    \item $m_{1}$: Number of Atoms: The count of heavy (non-hydrogen) atoms in the molecule, discretized into bins. Hydrogens are typically implicit in SMILES notation and are not counted. Range $(\text{min}_1, \text{max}_1) = (5, 30)$.
    \item $m_{2}$: Number of Bonds: The count of covalent bonds in the molecular graph between heavy atoms, similarly discretized into bins. Bonds to implicit hydrogen atoms are not counted. 
    Each bond is counted once, regardless of type (single and double bonds both count as 1). Range $(\text{min}_2, \text{max}_2) = (4, 32)$.
\end{itemize}

We utilize square $M \times M$ grid-shaped archives, where the index $b_i$ along dimension $i$ is:
\begin{equation}
        b_i = \left\lfloor \frac{m_i - \text{min}_i}{\text{max}_i - \text{min}_i} \times M \right\rfloor
\end{equation}
The value of $M$ depends on the binning scheme. There are two:
\begin{itemize}
    \item {Coarse}: $M = 10$: A $10 \times 10$ grid (100 bins). 
    \item {Fine}: $M = 20$: A $20 \times 20$ grid (400 bins).
\end{itemize}

This approach does not provide a one-to-one mapping between measure values and bins. 
Some adjacent measure values map to the same bin, even in the fine scheme. 
Molecules with measure values beyond the boundaries are assigned to the nearest bin. 

The coarse archive has fewer bins, storing fewer molecules and increasing competition within each niche. The possible benefit of fewer bins is that low fitness individuals are less likely to persist and be selected for mutation. The impact of archive granularity depends on how evolutionary stepping stones are distributed in measure space. A fine-grained archive might be better if important stepping stones lie across boundaries that are merged into a single bin in a coarse archive. This is one of several empirical questions we explore with our experiments. Also, note that with MOME, the Pareto fronts in each bin can have multiple molecules, so different molecules on the same front in a coarse archive may have measures that would result in separate placement in a fine-grained archive.

Regardless of the granularity, many bins in both archive types represent properties that are chemically impossible to fulfill, such as having more bonds than atoms. 







\subsection{SMILES String Mutations}
\label{sec:mutations}

We employ seven mutation operators~\cite{mashak:gecco2024, mashak:match2024} that stochastically modify canonical SMILES strings using RDKit~\cite{landrum:rdkit2010}, restricting mutations to \{C, N, O\} atoms with \{\texttt{-}, \texttt{=}\} bonds. Each mutation converts the parent SMILES to an editable molecular graph, applies the modification, and regenerates a canonical SMILES. Invalid results are discarded and retried up to 20 times before attempting a different mutation. The seven operators are:
\begin{itemize}
    \item {Change Bond Type}: Toggles single/double bonds respecting valence (e.g., \(\mathrm{C{=}C{-}N{-}O} \rightarrow \mathrm{C{-}C{-}N{-}O}\))
    \item {Insert Atom}: Breaks a bond and inserts C, N, or O \\(e.g., \(\mathrm{C{=}C{-}N{-}O} \rightarrow \mathrm{C{=}C{-}O{-}N{-}O}\))
    \item {Add Branch}: Attaches a new atom to an existing one \\(e.g., \(\mathrm{C{=}C{-}N{-}O} \rightarrow \mathrm{C{=}C({-}O){-}N{-}O}\))
    \item {Delete Atom}: Removes an atom and reconnects the graph (e.g., \(\mathrm{C{=}C{-}N{-}O} \rightarrow \mathrm{C{=}C{-}O}\))
    \item {Change Atom Type}: Substitutes atom element \\ (e.g., \(\mathrm{C{=}C{-}N{-}O} \rightarrow \mathrm{C{=}C{-}N{-}C}\))
    \item {Add Ring}: Connects non-adjacent atoms to form a cycle \\ (e.g., \(\mathrm{C{=}C{-}N{-}O} \rightarrow \mathrm{C1{=}C{-}N{-}O1}\))
    \item {Delete Ring Bond}: Opens a ring by removing one bond \\ (e.g., \(\mathrm{C1{=}C{-}N{-}O1} \rightarrow \mathrm{C{=}C{-}N{-}O}\))
\end{itemize}

Each child (except in simulated annealing) receives one to three randomly selected mutations. Simulated annealing generates one child per mutation type.



\section{Algorithmic Settings}
\label{sec:settings}

Each algorithm evaluates approximately 2,000 molecules. 
Simulated annealing starts from one molecule, applies all seven mutations per iteration for 285 iterations (1,996 evaluations), with exponential cooling from $T_{\text{initial}} = 100$ to $T_{\text{min}} = 0.01$ at rate $\alpha = 0.95$.

NSGA-II and $(\mu+\lambda)$ use $\mu = \lambda = 20$ over 100 generations (2,020 evaluations) with tournament selection ($k=3$). 

MAP-Elites and MOME seed archives with 50 random molecules, then mutate randomly selected archive members for 100 generations of 20 evaluations each (2,050 total). 


All experiments use 20 random seeds. Initial populations are generated from common scaffolds (e.g., C, C=C, C-C-N) with multiple random mutations applied to create valid, unique molecules. Only C, N, O, and H atoms are permitted. Single and double bonds are allowed; triple bonds are forbidden. Molecules must contain 5-30 heavy atoms (non-hydrogen) and must form a single connected component (no disconnected fragments).

\section{Results}

Algorithms are compared in terms of the following results produced by each of their 20 distinct runs using different random seeds.
In general, results across runs do not follow a normal distribution and have very high performing outliers, 
so we favor the use of median scores to compare typical algorithm performance.

\subsection{Median Best Objective Scores}

\begin{figure*}[t]
\centering
\begin{subfigure}[t]{0.25\textwidth}
    \centering
    \includegraphics[width=\linewidth]{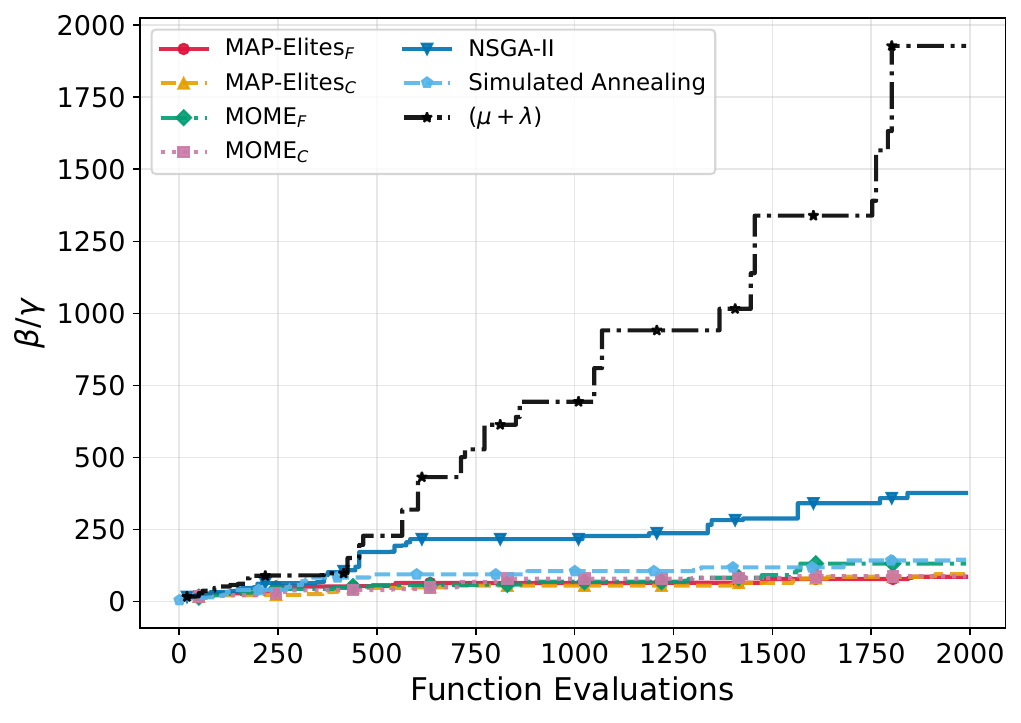}
    \caption{\small Maximize $\beta/\gamma$}
    \label{fig:beta_gamma}
\end{subfigure}\hfill
\begin{subfigure}[t]{0.25\textwidth}
    \centering
    \includegraphics[width=\linewidth]{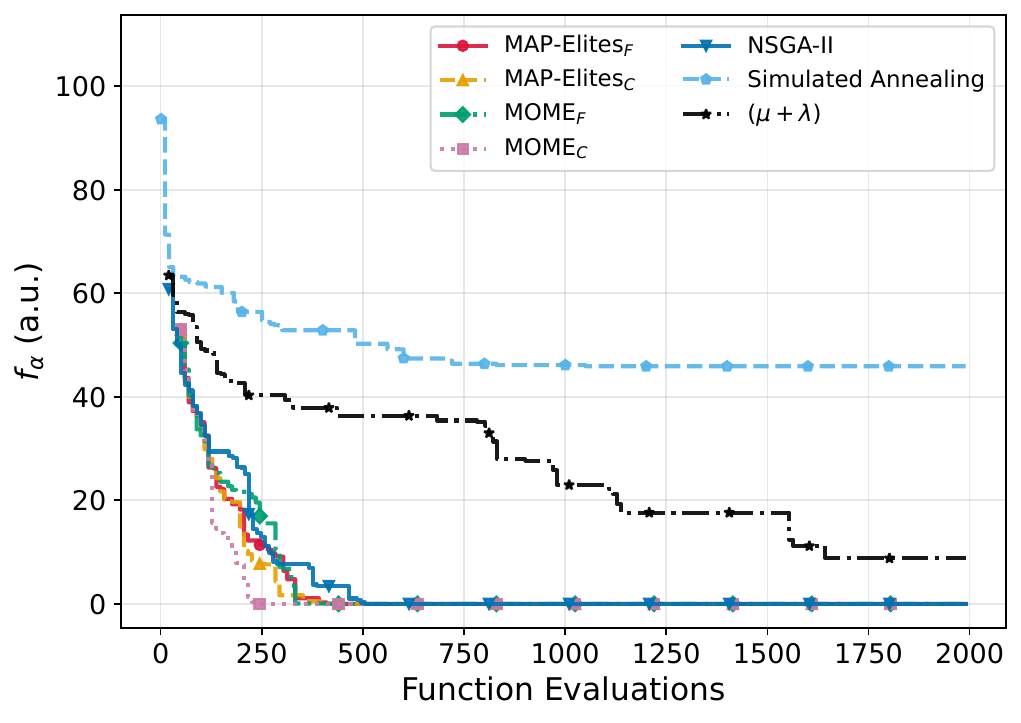}
    \caption{\small Minimize $f_{\alpha}$}
    \label{fig:lin_pol}
\end{subfigure}\hfill
\begin{subfigure}[t]{0.25\textwidth}
    \centering
    \includegraphics[width=\linewidth]{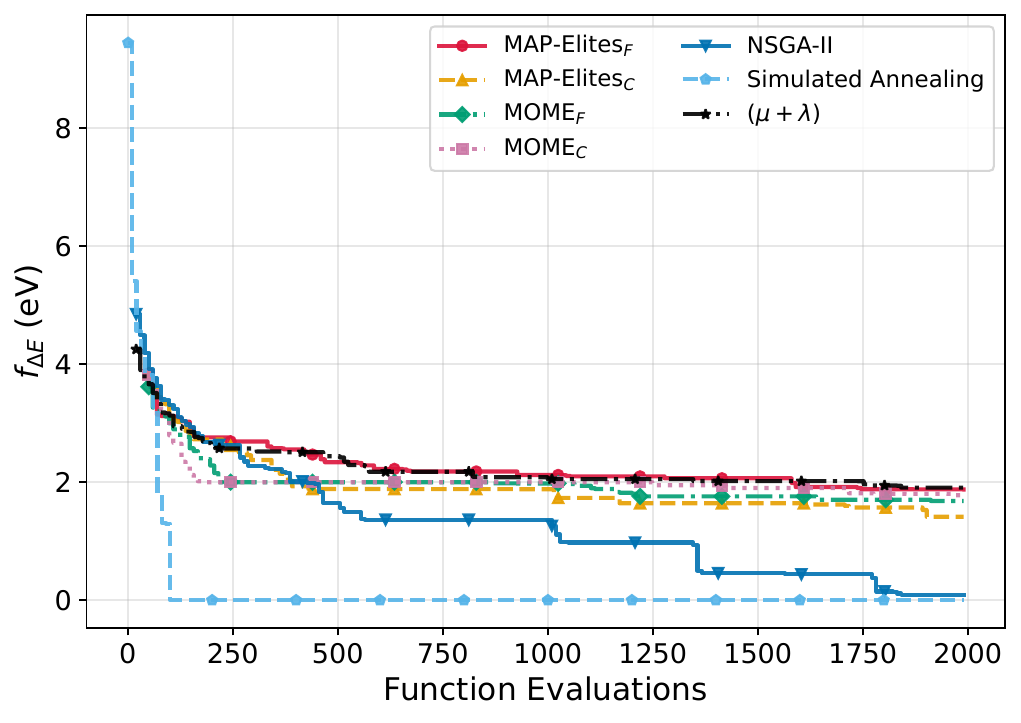}
    \caption{\small Minimize $f_{\Delta E}$}
    \label{fig:homo_lumo}
\end{subfigure}\hfill
\begin{subfigure}[t]{0.25\textwidth}
    \centering
    \includegraphics[width=\linewidth]{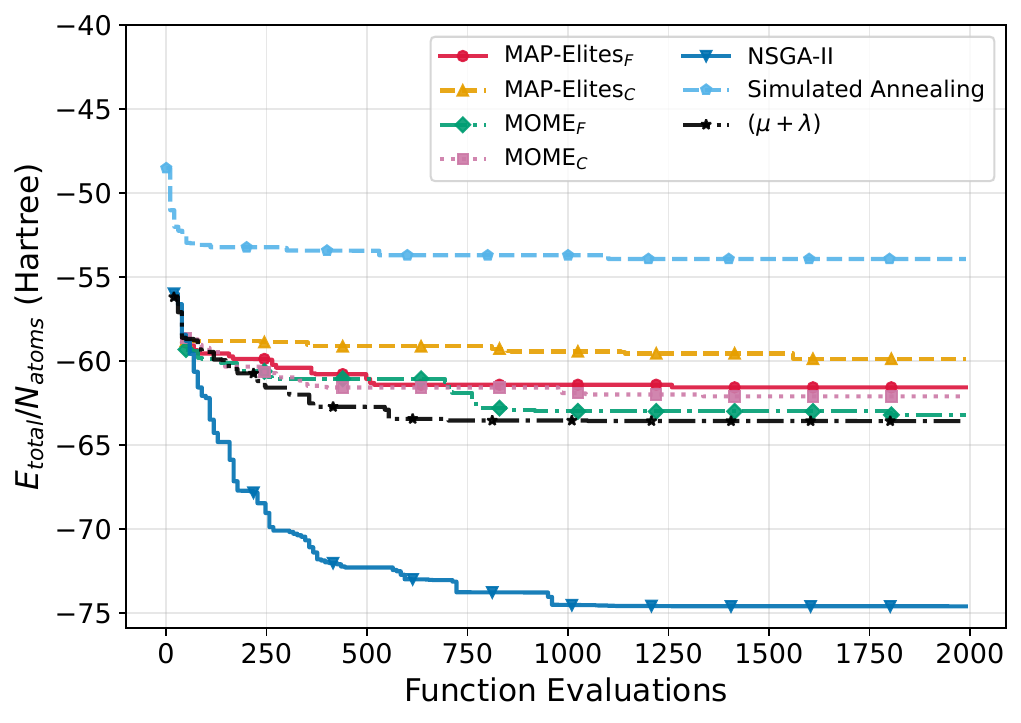}
    \caption{\small Minimize $E_{total}/N_{atoms}$}
    \label{fig:energy_per_atom}
\end{subfigure}

\caption{\small Median Best Objective Scores Across 20 Runs of Each Algorithm:
(\subref{fig:beta_gamma}) Median first-to-second hyperpolarizability ratio.
High values are better, so
$(\mu+\lambda)$ outperforms all others by a large margin, including other single-objective methods,
though NSGA-II is clearly second-best.
(\subref{fig:lin_pol}) Median linear polarizability range deviation.
All but simulated annealing and $(\mu+\lambda)$ quickly reach a perfect minimal score of 0,
though $(\mu+\lambda)$ at least gets close.
(\subref{fig:homo_lumo}) Median $f_{\Delta E}$ range deviation.
NSGA-II and simulated annealing tie for best with perfect minimal scores of 0,
which the other algorithms do not reach. The single-objective methods were not
aware of this objective, so their poorer performance is not surprising,
but MOME's performance is slightly disappointing. However, these are only median scores;
some MOME runs reach the perfect score, but less than half.
(\subref{fig:energy_per_atom}) Median energy per atom.
NSGA-II is clearly the best at minimizing this objective, with most other methods
clustering closer together, including the single-objective methods that were unaware of this objective.
Only simulated annealing is exceptionally poor.}
\label{fig:objectives}
\Description[Median Best Objective Scores Across 20 Runs of Each Algorithm]{TODO}
\end{figure*}



Across all
seeds of each algorithm, the median of the \emph{best} objective scores ever encountered in each objective
are in Figure \ref{fig:objectives}. 
The best objective score after a certain number of evaluations
in a given run is either the highest or lowest score,
depending on the objective, across all molecules produced up to that point.

The incredibly high $\beta/\gamma$ scores of plain $(\mu+\lambda)$ selection stand out,
but when combined with poorer scores in other objectives,
these molecules actually prove to be so unstable as to be practically worthless.
The unfortunate conclusion is that optimizing exclusively based on hyperpolarizability scores
allows evolution to cheat in a way that bypasses the intended application of these molecules.
The push for diversity in MAP-Elites is likely what helps it avoid this trap, despite also being a single-objective method.
Simulated annealing does surprisingly well in $f_{\Delta E}$ score, despite this not being the objective it was explicitly optimizing, but at the cost of $f_{\alpha}$ and $E_{total}/N_{atoms}$. The high $E_{total}/N_{atoms}$ indicates severe instability.

The next highest $\beta/\gamma$ scores come from NSGA-II, which also achieves top performance in the other
three objectives, tying for best in $f_{\alpha}$, 
being the only evolutionary method with an optimal median for $f_{\Delta E}$, and being the clear best in $E_{total}/N_{atoms}$.
These results make sense given the MO nature of NSGA-II.




\subsection{Median Global Hypervolume}
\label{sec:medianglobalhv}

\begin{figure*}[t]
\centering
\begin{subfigure}{0.45\textwidth}
    \centering
    \includegraphics[width=\linewidth]{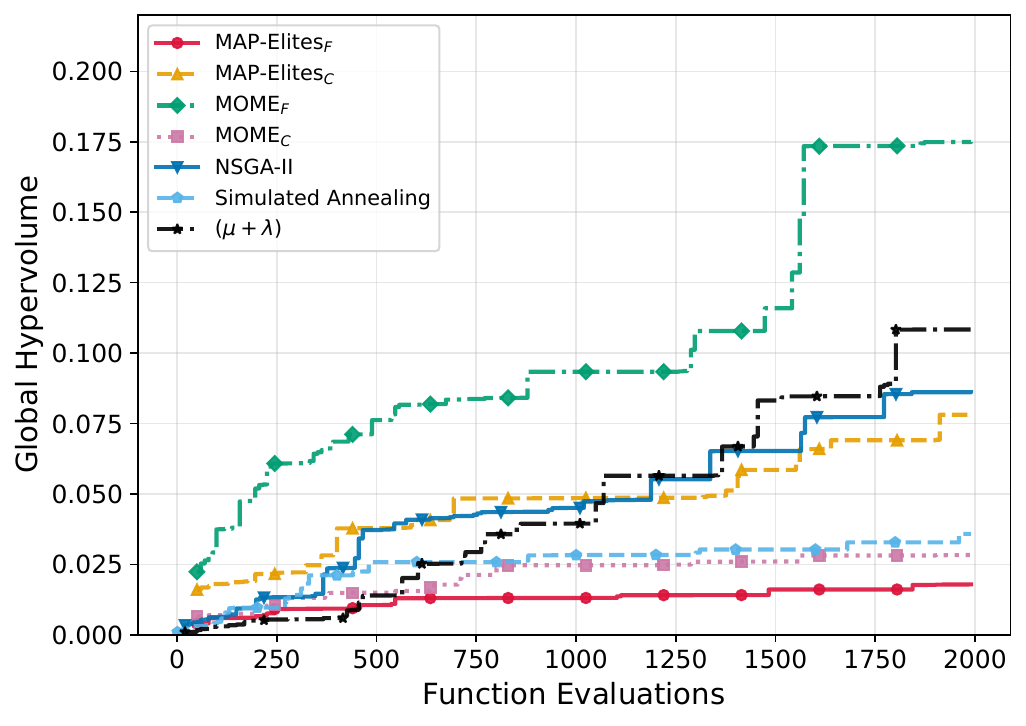}
    \caption{Across Evaluations}
    \label{fig:global_hv_eval}
\end{subfigure}\hfill
\begin{subfigure}{0.45\textwidth}
    \centering
    \includegraphics[width=\linewidth]{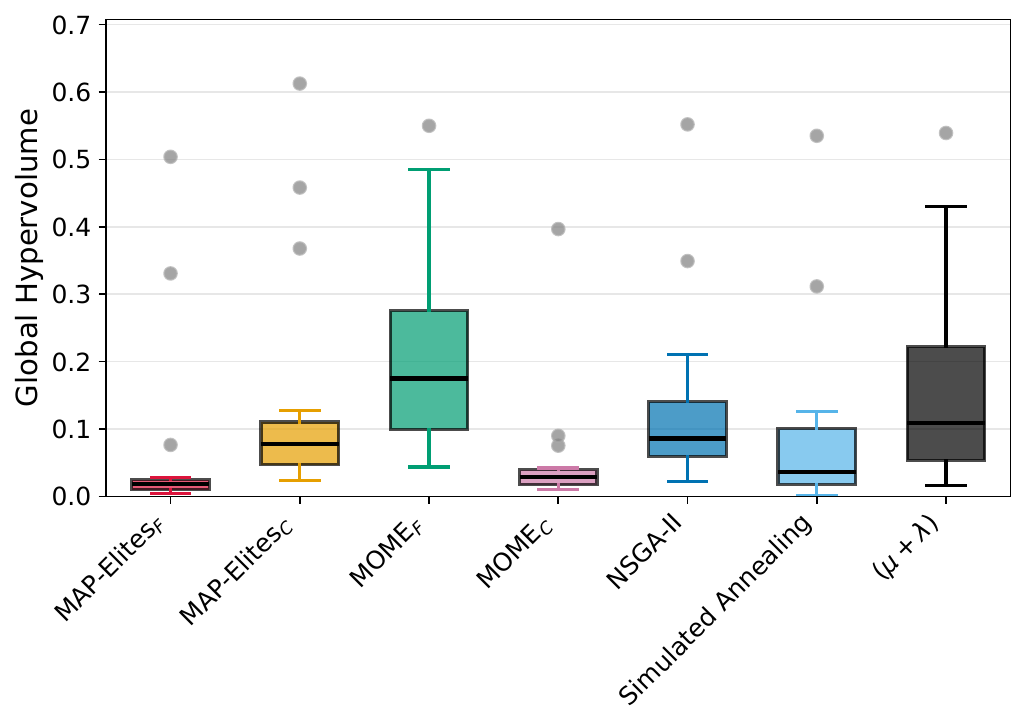}
    \caption{Final Pareto Fronts}
    \label{fig:global_hv_final}
\end{subfigure}

\caption{\small Global Hypervolume Scores Across 20 Runs of Each Algorithm:
(\subref{fig:global_hv_eval}) Median hypervolume scores for each algorithm across function evaluations.
$\text{MOME}_{F}$ is the best, followed by a cluster of $(\mu+\lambda)$, NSGA-II, and $\text{MAP-Elites}_{C}$,
before algorithms start to bunch together near the bottom.
(\subref{fig:global_hv_final}) Box-and-whisker plots of hypervolume scores for final Pareto fronts.
The lower quartile, median, and upper quartile are the lower boundary, center line, and upper boundary of each box respectively, and the whiskers denote the furthest points within $1.5IQR$ of the nearest quartile, where $IQR$ is the interquartile range. Points outside of the whiskers are outliers, of which there are many.
However, $\text{MOME}_{F}$ and $(\mu+\lambda)$ both have high upper quartiles, and spread more across the range of higher scores.}
\label{fig:global_hv}
\Description[Median Global Hypervolume Scores Across 20 Runs of Each Algorithm]{TODO}
\end{figure*}


Regardless of whether evolution is single- or multi-objective, 
a Pareto front can be computed over all molecules with our four objectives. Therefore, we calculate a Pareto
front across all molecules generated so far in each run, and use it to calculate the hypervolume using \texttt{pymoo} \cite{blank:ieee2020}.
Hypervolume (HV) measures the region (Lebesgue measure) dominated by the Pareto front 
with respect to a reference point dominated by all points \cite{zitzler:tec2003}.
HV is a single metric measuring the quality of trade-offs in the front.

However, hypervolume is sensitive to 
differences in objective scale and to extreme outliers \cite{zitzler:tec2003},
making normalization of objective scores necessary.
Our quantum chemistry calculations sometimes produced extremes 
that were impossible, which needed to be discarded (see Section \ref{sec:discussion}).
Based on analysis of what evolution discovered, and informed by knowledge of
what is reasonable both chemically and physically, 
we normalized each objective score based on the following ranges:
\begin{itemize}
\item $\beta/\gamma$: $[0, 9419]$ (unitless)
\item $f_{\alpha}$: $[0, 440]$ (a.u.)
\item $f_{\Delta E}$: $[0, 16]$ (eV)
\item $E_{total}/N_{atoms}$: $[-75, 0]$ (Hartree)
\end{itemize}
These values scaled scores to the range $[0,1]$ before performing the HV calculation with a reference point of $\vec{0}$.


Figure \ref{fig:global_hv_eval} shows median hypervolume scores across evaluations. We call these \emph{global} hypervolumes to distinguish them from scores associated with MOQD (Section~\ref{sec:moqd}).
$\text{MOME}_{F}$ performs best by this metric, which makes sense given that it is explicitly multiobjective and maintains diverse niches, which helps with coverage of objective space. 
In contrast, the relatively poor performance of $\text{MOME}_{C}$ is surprising. The $\text{MOME}_{C}$ results suggest an overall quality penalty from having
fewer bins to support more fine-grained niches. 

Below $\text{MOME}_{F}$ are methods roughly tied for second place in terms of median HV: $(\mu+\lambda)$, NSGA-II, and $\text{MAP-Elites}_{C}$. NSGA-II's performance makes sense given that MO algorithms are meant to increase HV. 
The $(\mu+\lambda)$ result hides the fact that this HV score is primarily due to high $\beta/\gamma$, and not balanced across objectives.

The high performance of $\text{MAP-Elites}_{C}$ contrasts with MOME: $\text{MAP-Elites}_{C}$ is far superior to $\text{MAP-Elites}_{F}$ (worst global HV), but the coarse vs.\ fine relationship is opposite with MOME.
A fine-grained archive is beneficial to MOME, but a detriment to MAP-Elites,
likely due to the specific objectives and measures used. Changes in atom or bond count 
are more likely to have a direct impact on 
HOMO-LUMO gap, energy per atom, and linear polarizability,
which are tied to
the objectives that MAP-Elites does not directly optimize.
Therefore, each $\text{MAP-Elites}_{C}$ 
bin forces molecules with differing atom and bond counts to compete on $\beta/\gamma$, while allowing variation in unoptimized objectives to persist among high-performing solutions. This can increase global hypervolume by retaining non-dominated trade-offs that are a byproduct of optimizing $\beta/\gamma$. In contrast, coarse MOME archives increase domination pressure in each bin, causing trade-offs between structurally dissimilar molecules to cut off possible stepping stones, reducing the diversity
of molecules for further exploration.

Figure \ref{fig:global_hv_final} focuses only on results at the end of evolution, but shows spread across all 20 runs
of each algorithm with box-and-whisker plots. There are many high outliers. The only algorithms whose scores
are consistently in a higher range are $\text{MOME}_{F}$ and $(\mu+\lambda)$. The high performance of $\text{MOME}_{F}$ makes
sense for the reasons described above, whereas the high HV scores for $(\mu+\lambda)$ depend primarily on
the contribution of its high $\beta/\gamma$ scores, as shown previously. Even though hypervolume is a measure of
space dominated by \emph{all} objectives, the $\beta/\gamma$ scores for $(\mu+\lambda)$ are so high that they compensate
for poor scores in the other objectives.


\subsection{Fine and Coarse Median Archive Bin Count}


\begin{figure*}[t]
\centering
\begin{subfigure}[t]{0.25\textwidth}
    \centering
    \includegraphics[width=\linewidth]{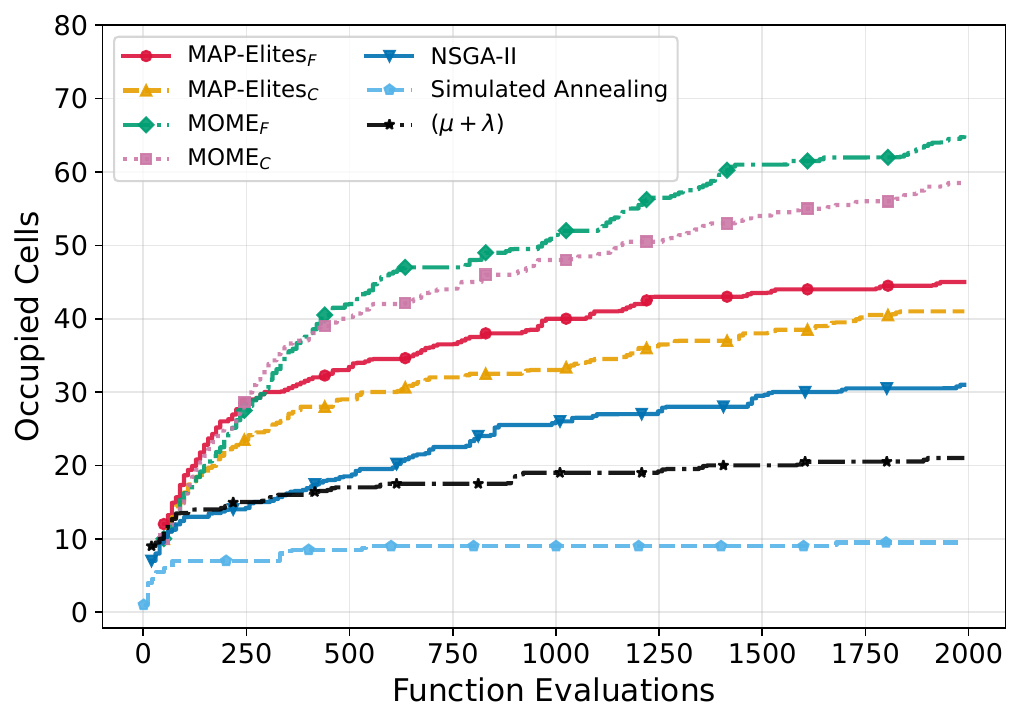}
    \caption{\small $\text{Count}_{F}$}
    \label{fig:count_f}
\end{subfigure}\hfill
\begin{subfigure}[t]{0.25\textwidth}
    \centering
    \includegraphics[width=\linewidth]{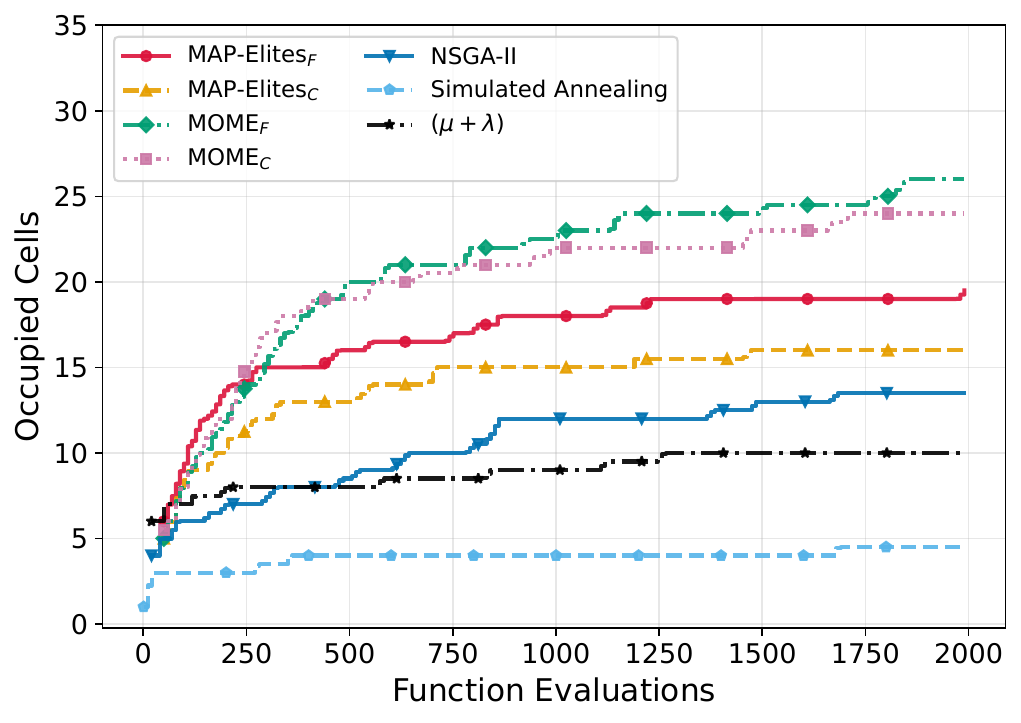}
    \caption{\small $\text{Count}_{C}$}
    \label{fig:count_c}
\end{subfigure}\hfill
\begin{subfigure}[t]{0.25\textwidth}
    \centering
    \includegraphics[width=\linewidth]{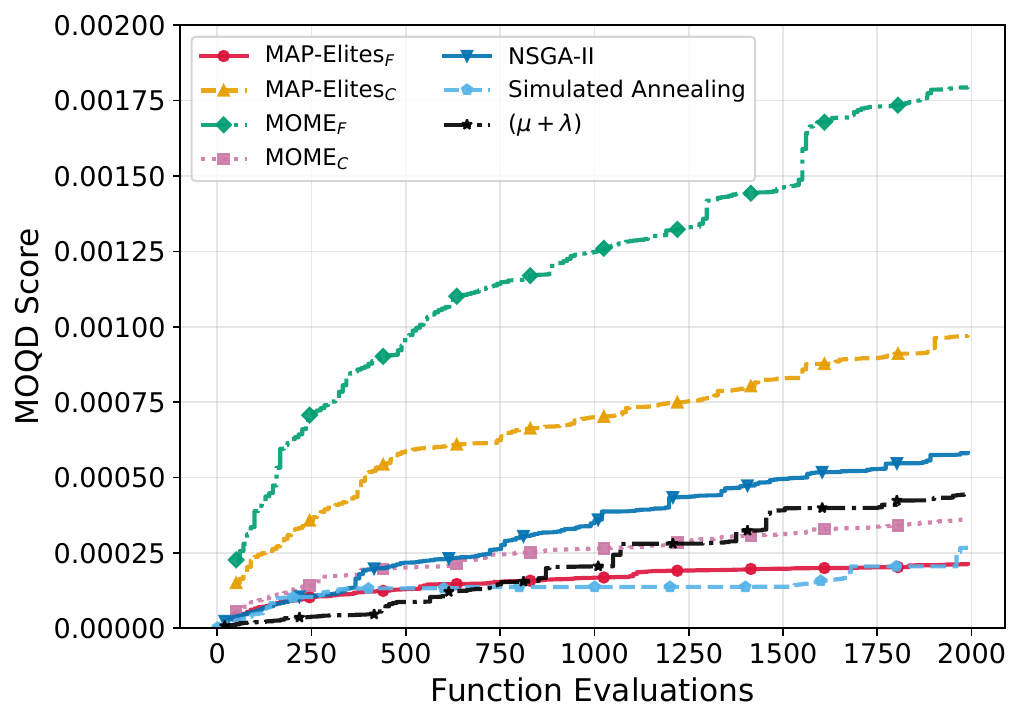}
    \caption{\small $\text{MOQD}_{F}$}
    \label{fig:moqd_f}
\end{subfigure}\hfill
\begin{subfigure}[t]{0.25\textwidth}
    \centering
    \includegraphics[width=\linewidth]{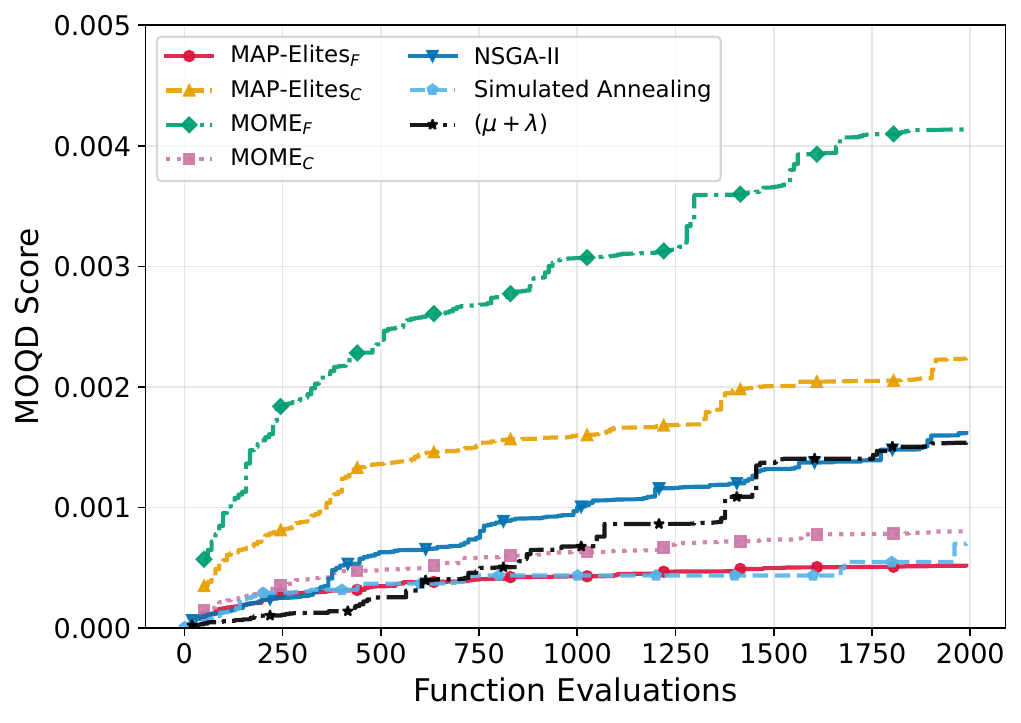}
    \caption{\small $\text{MOQD}_{C}$}
    \label{fig:moqd_c}
\end{subfigure}

\caption{\small Fine-grained and Coarse Archive Median Scores Across 20 Runs of Each Algorithm:
(\subref{fig:count_f}) Median bin count with fine-grained binning.
Unsurprisingly, QD methods fill more bins than non-QD approaches, and simulated annealing
performs the worst. Fine-grained QD methods occupy slightly more bins than their coarse counterparts. 
(\subref{fig:count_c}) Median bin count with coarse binning.
The scale is different with a coarse archive, but results are qualitatively identical
to the fine-grained archive results. Interestingly, fine-grained QD methods still occupy more bins
than their coarse counterparts.
(\subref{fig:moqd_f}) Median MOQD with fine-grained binning.
$\text{MOME}_{F}$ is clearly superior, followed distantly by $\text{MAP-Elites}_{C}$, then NSGA-II, before the rest cluster more tightly together.
(\subref{fig:moqd_c}) Median MOQD with coarse binning.
Qualitatively similar to the fine-grained MOQD results, except that
$(\mu+\lambda)$ demonstrates a significant jump in MOQD score near the end of evolution that ties it with NSGA-II.}
\label{fig:fine_coarse_count_moqd}
\Description[Fine-grained and Coarse Archive Median Scores Across 20 Runs of Each Algorithm]{TODO}
\end{figure*}

With QD methods, it is desirable to know how much of the measure space is covered by solutions.
For algorithms with archives, we simply count the number of occupied bins, but the comparison
is more complicated when different methods use different binning schemes, or do not use an archive at all.

For the sake of fair comparison, we present both the number of bins that would be occupied 
in a fine-grained binning scheme $\text{Count}_{F}$ (Figure \ref{fig:count_f}), 
and the number of bins that would be occupied in
a coarse binning scheme $\text{Count}_{C}$ (Figure \ref{fig:count_c}). 
Note that $\text{Count}_{F}$ can be calculated for
QD methods that evolve with a coarse archive and vice-versa. These quantities are also calculated for
algorithms that do not use an archive, using the stream of generated molecules
to create the archive after the algorithm is finished.
Each Count metric
is the number of occupied bins in an archive of the 
appropriate type after a given number of molecule evaluations.

Fine and coarse archive counts exhibit differences in scale but are
not qualitatively distinct. Each exhibits the same ordering of algorithms
and similar spacing between each.
QD/MOQD methods explicitly support diversity, and thus occupy more bins than 
non-QD algorithms. Simulated annealing is especially poor in this metric,
since it does not even use a population, hindering diversity.


Interestingly, fine-grained versions of MOME and MAP-Elites
fill more bins in the coarse archive than the coarse versions, but the
margins are small. Similarly, MOME with a given binning scheme occupies
slightly more bins than MAP-Elites with that same binning scheme in both
the coarse and fine cases. 

\subsection{Fine and Coarse Median QD Score}

\begin{figure*}[t]
\centering
\begin{subfigure}[t]{0.25\textwidth}
    \centering
    \includegraphics[width=\linewidth]{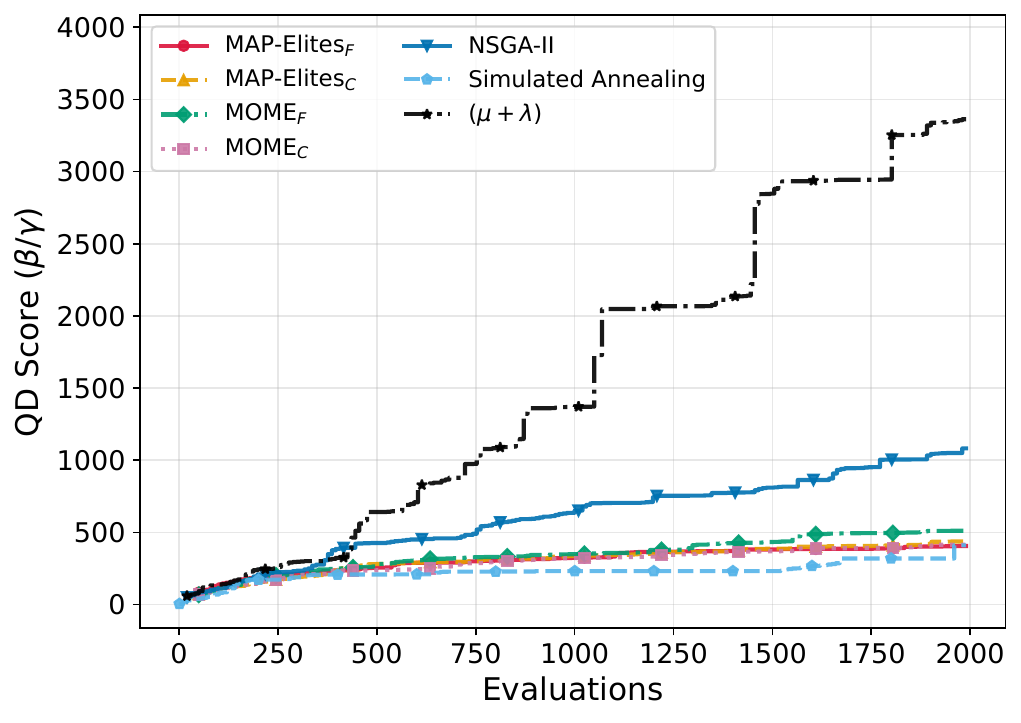}
    \caption{\small $(\beta/\gamma)_{F}$}
    \label{fig:beta_gamma_fine_qd}
\end{subfigure}\hfill
\begin{subfigure}[t]{0.25\textwidth}
    \centering
    \includegraphics[width=\linewidth]{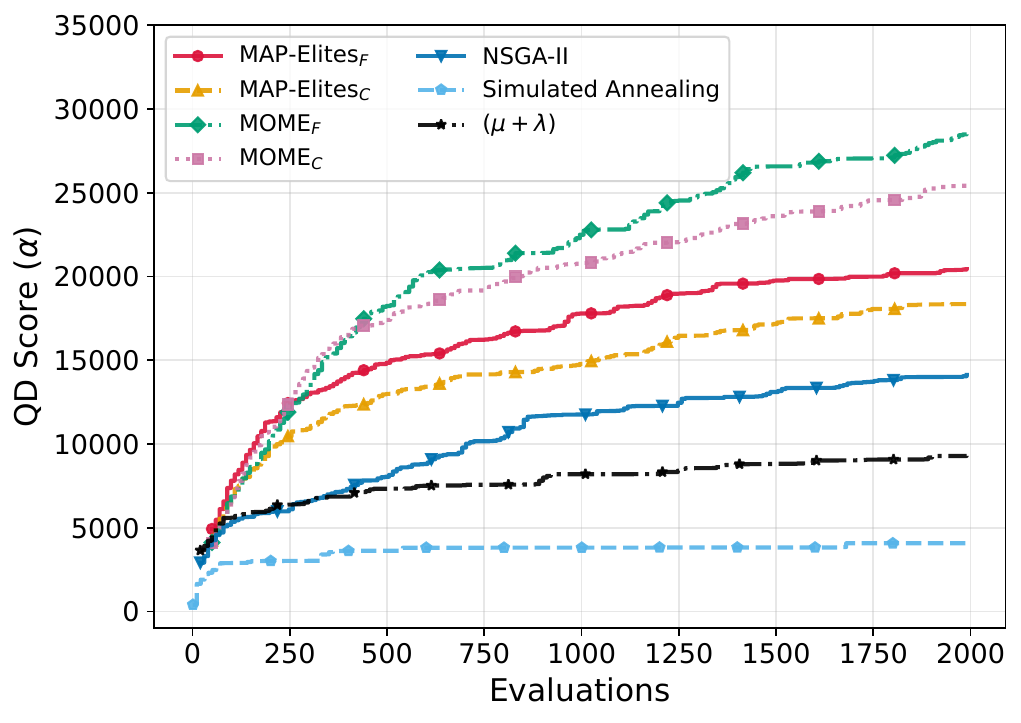}
    \caption{\small $(f_{\alpha})_{F}$}
    \label{fig:lin_pol_fine_qd}
\end{subfigure}\hfill
\begin{subfigure}[t]{0.25\textwidth}
    \centering
    \includegraphics[width=\linewidth]{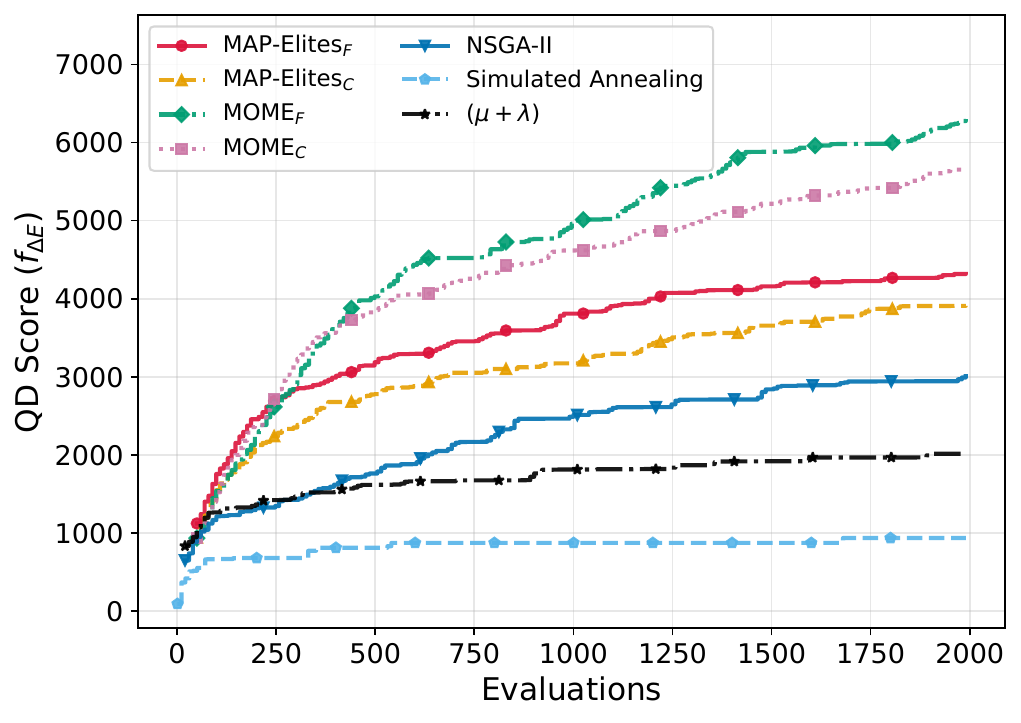}
    \caption{\small $(f_{\Delta E})_{F}$}
    \label{fig:homo_lumo_fine_qd}
\end{subfigure}\hfill
\begin{subfigure}[t]{0.25\textwidth}
    \centering
    \includegraphics[width=\linewidth]{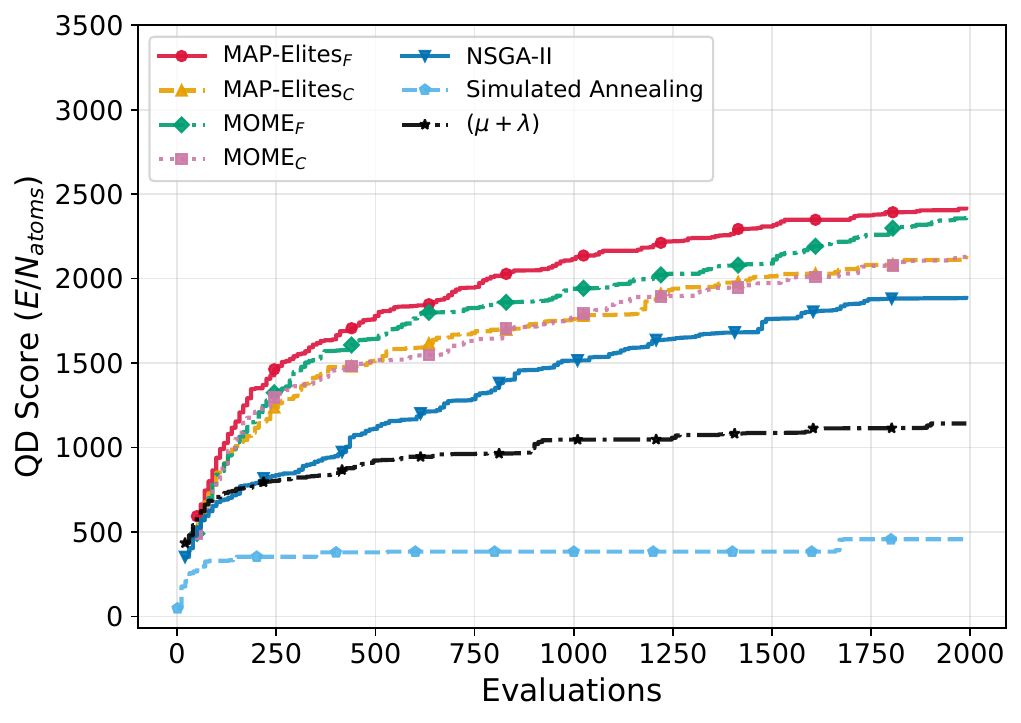}
    \caption{\small $(E_{total}/N_{atoms})_{F}$}
    \label{fig:energy_per_atom_fine_qd}
\end{subfigure}

\caption{\small Fine-grained Archive Median QD Scores by Objective Across 20 Runs of Each Algorithm:
(\subref{fig:beta_gamma_fine_qd}) Median QD for first-to-second hyperpolarizability ratio using fine-grained binning.
Qualitatively similar to raw objective scores for $\beta/\gamma$ (Figure \ref{fig:beta_gamma_fine_qd}),
with strong performance by $(\mu+\lambda)$ and NSGA-II in second place.
(\subref{fig:lin_pol_fine_qd}) Median QD for linear polarizability range deviation using fine-grained binning.
Both MOME approaches perform the best, with MAP-Elites approaches beneath them, and NSGA-II
trailing close behind, beating $(\mu+\lambda)$ and simulated annealing.
(\subref{fig:homo_lumo_fine_qd}) Median QD for HOMO-LUMO gap range deviation using fine-grained binning.
Qualitatively similar to the $(f_{\alpha})_{F}$ results.
(\subref{fig:energy_per_atom_fine_qd}) Median QD for energy per atom using fine-grained binning.
The MOQD and QD methods are more tightly clustered with NSGA-II near the top, but $(\mu+\lambda)$
is still far behind and simulated annealing is far below that.}
\label{fig:qd_f}
\Description[Fine-grained Archive Median QD Scores by Objective Across 20 Runs of Each Algorithm]{TODO}
\end{figure*}


Single-objective QD algorithms use QD score \cite{pugh:gecco2015} to measure success.
QD score is the sum of scores in the one objective across occupied bins in an archive,
meaning that QD score increases when more bins are occupied and when objective scores in occupied bins improve.
Generating more and better solutions increases QD score.

Fine and coarse archives are used to 
calculate QD score according to each
objective. 
We add the stream of generated molecules to an archive, and only allow a molecule in an occupied bin to be replaced
if the new molecule has a better score in the objective of interest. 

Figure \ref{fig:qd_f} shows QD scores associated with each objective for fine-grained archives. Results for
coarse archives are qualitatively identical, so are excluded. 
QD scores for $\beta/\gamma$ (Figure~\ref{fig:beta_gamma_fine_qd}) are qualitatively similar to the raw $\beta/\gamma$ scores (Figure~\ref{fig:beta_gamma}). QD scores based on the remaining objectives demonstrate the strengths of MOME:
fine-grained and coarse versions excel in these objectives.

Though worse than MOME for $(f_{\alpha})_{F}$ and $(f_{\Delta E})_{F}$, 
MAP-Elites earns surprisingly high QD scores in 
these objectives, which it was not aware of, 
and $\text{MAP-Elites}_{F}$ achieves the best scores in
$(E_{total}/N_{atoms})_{F}$.
Even though NSGA-II
explicitly optimizes all objectives, MAP-Elites does better in these objectives by encouraging
diversity with $\beta/\gamma$ as the only objective. In contrast, $(\mu+\lambda)$ and simulated annealing perform poorly in 
objectives they were unaware off, due to lack of awareness and lack of pressure to diversify.

\subsection{Median Multiobjective QD Score}
\label{sec:moqd}


The paper introducing MOME also introduced the MOQD score, or multiobjective QD score \cite{pierrot:gecco2022}. Recall that
MOME bins can contain multiple solutions representing a Pareto front across objectives. Because HV 
represents the quality of a Pareto front, the quality of a MOME bin is the
hypervolume of that bin, and MOQD score is the sum of all hypervolumes across all occupied bins
in the archive.

Although this metric was developed for MOME, it can be applied to any algorithm given a stream of molecules.
Now molecules are only added to occupied bins if they are non-dominated. Thus, even for single-objective
QD algorithms and archive-free methods, $\text{MOQD}_{F}$ (Figure \ref{fig:moqd_f}) 
and $\text{MOQD}_{C}$ (Figure \ref{fig:moqd_c}) can be calculated.

$\text{MOME}_{F}$ is vastly superior in both fine-grained and coarse archives,
which is consistent with global HV results. 
$\text{MAP-Elites}_{C}$ is second best, being far below $\text{MOME}_{F}$. Next comes NSGA-II followed by the rest,
though in the coarse case $(\mu+\lambda)$ actually catches up to
NSGA-II before the end of evolution.
Still, the MOQD scores of $(\mu+\lambda)$ are much lower than its global hypervolume scores, indicating that its high HV
depends on a cluster of structurally similar molecules that do not occupy much of measure space (Section \ref{sec:megaHVheatmaps}).

\subsection{Mega Archive Hypervolume Heatmaps}
\label{sec:megaHVheatmaps}

\begin{figure*}[t]
\centering
\begin{subfigure}[t]{0.14\textwidth}
    \centering
    \includegraphics[width=\linewidth]{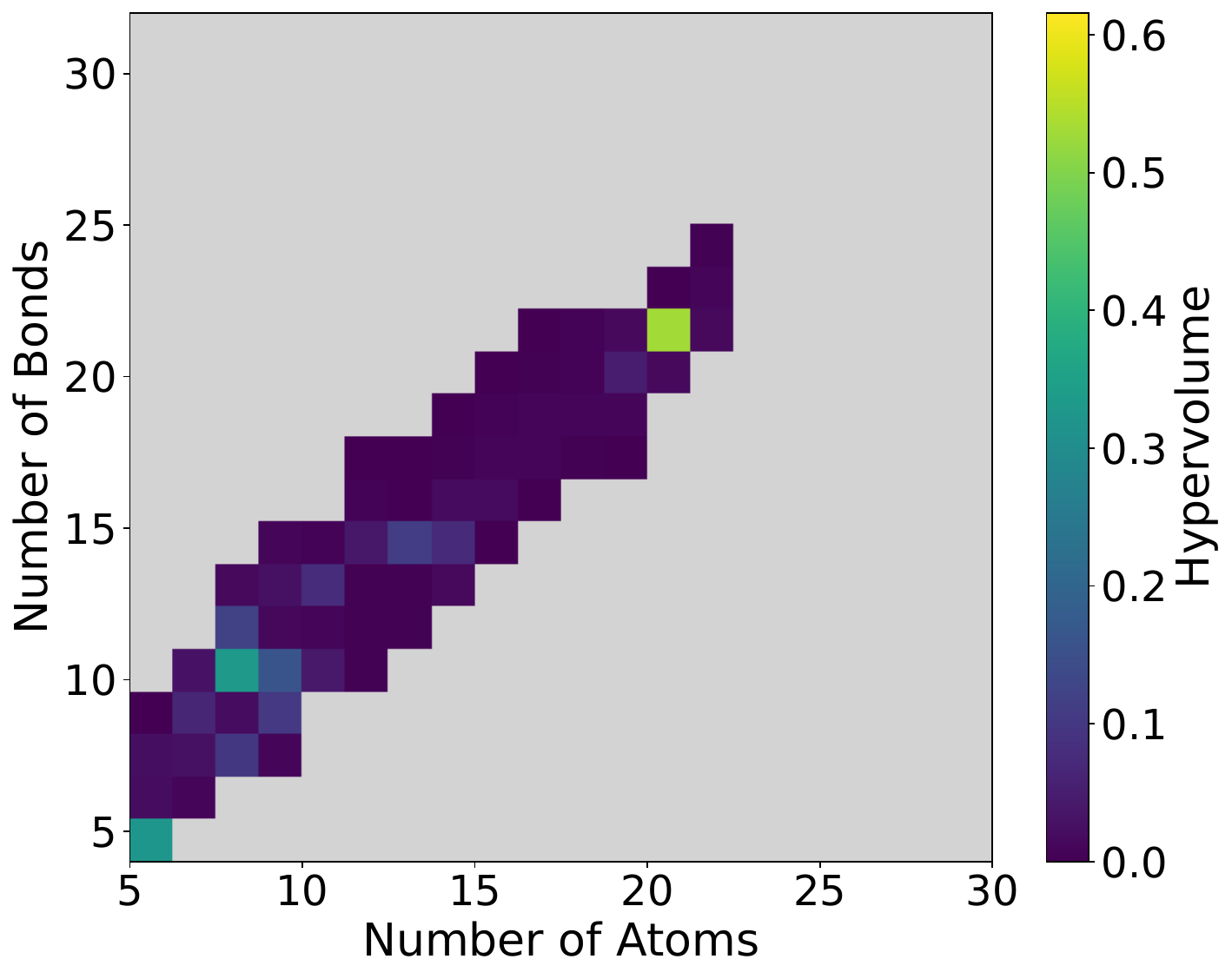}
    \caption{\small SA}
    \label{fig:f_sa_hv_heat}
\end{subfigure}\hfill
\begin{subfigure}[t]{0.14\textwidth}
    \centering
    \includegraphics[width=\linewidth]{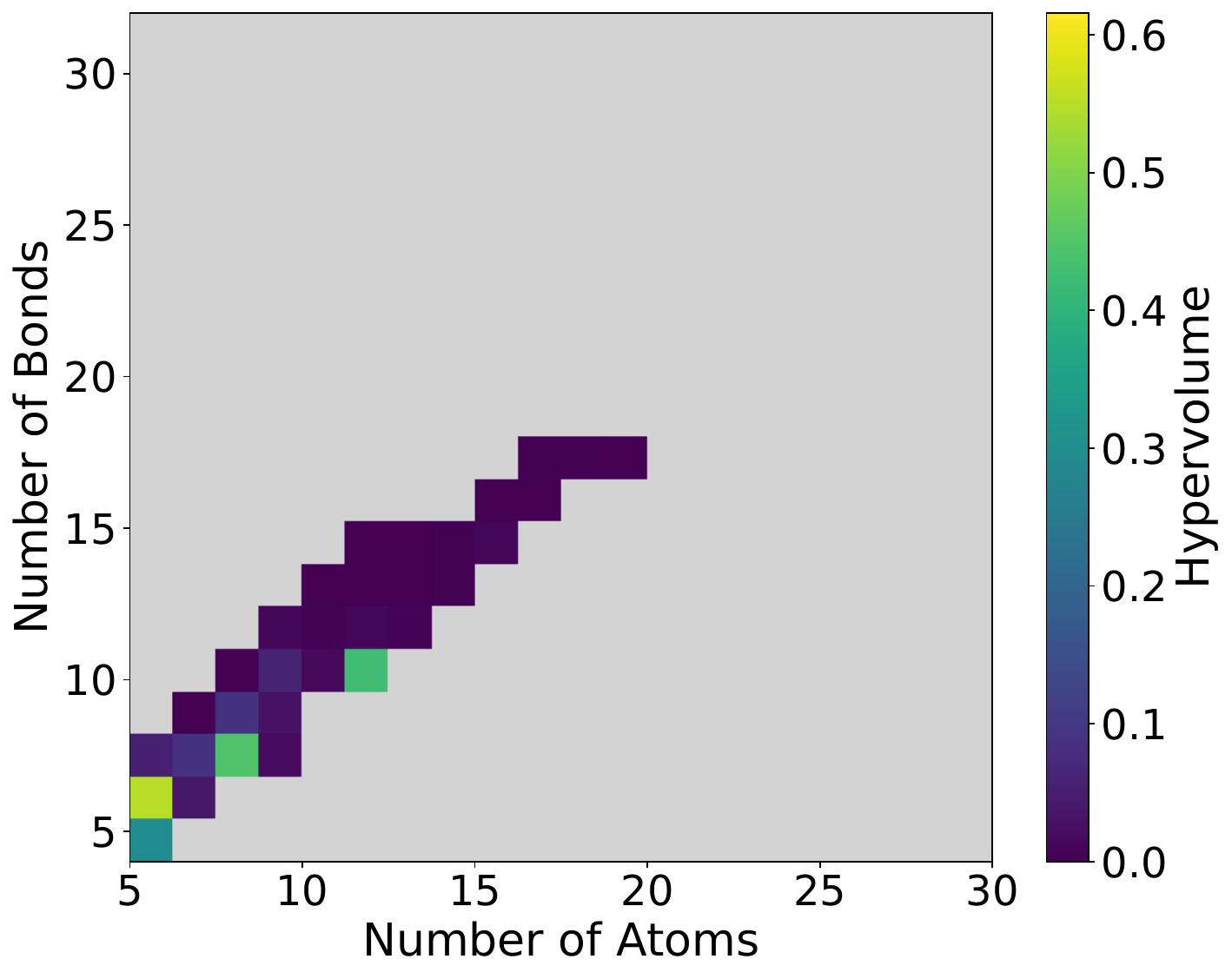}
    \caption{\small $(\mu+\lambda)$}
    \label{fig:f_ml_hv_heat}
\end{subfigure}\hfill
\begin{subfigure}[t]{0.14\textwidth}
    \centering
    \includegraphics[width=\linewidth]{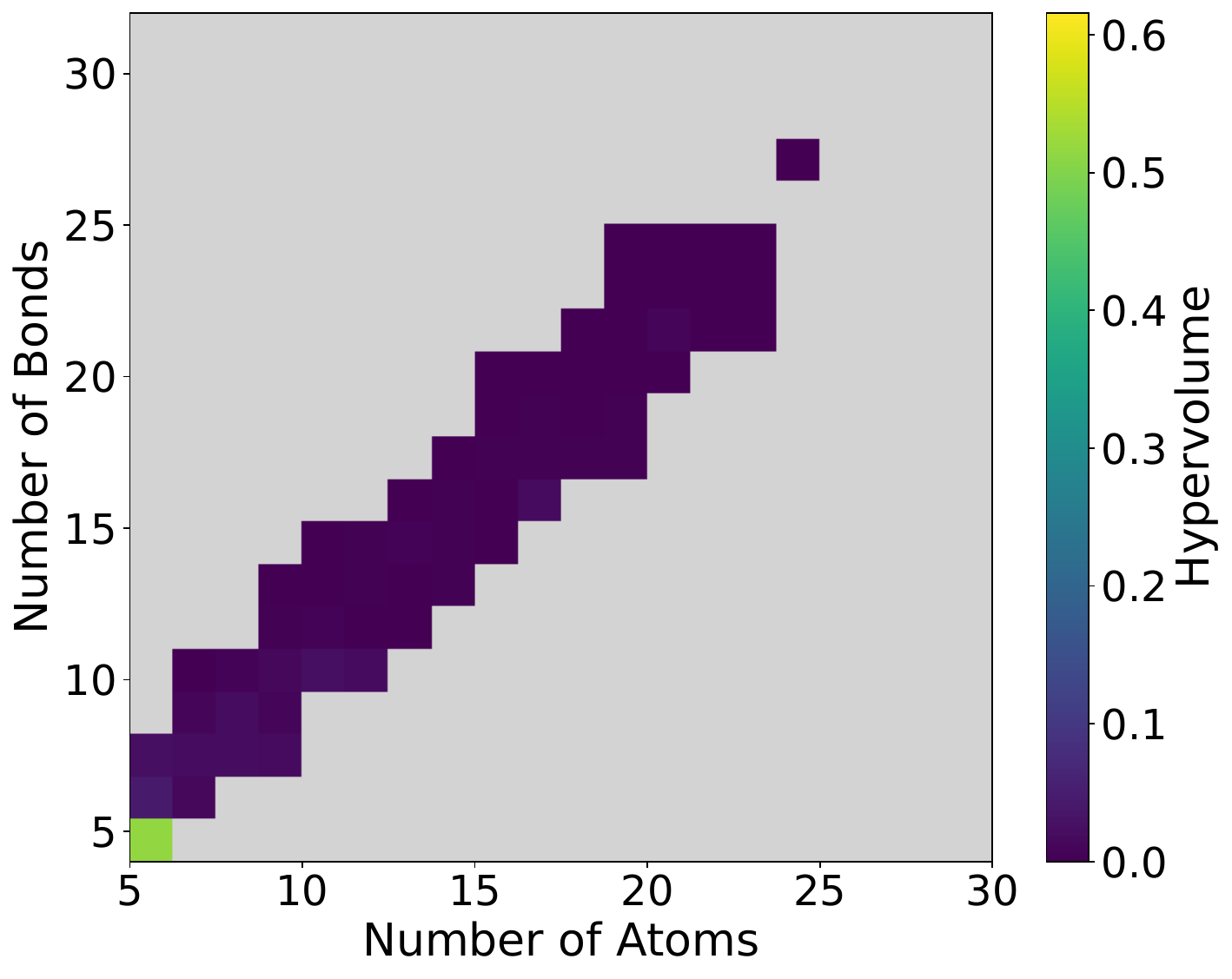}
    \caption{\small $\text{MAP-Elites}_{F}$}
    \label{fig:f_me_f_hv_heat}
\end{subfigure}\hfill
\begin{subfigure}[t]{0.14\textwidth}
    \centering
    \includegraphics[width=\linewidth]{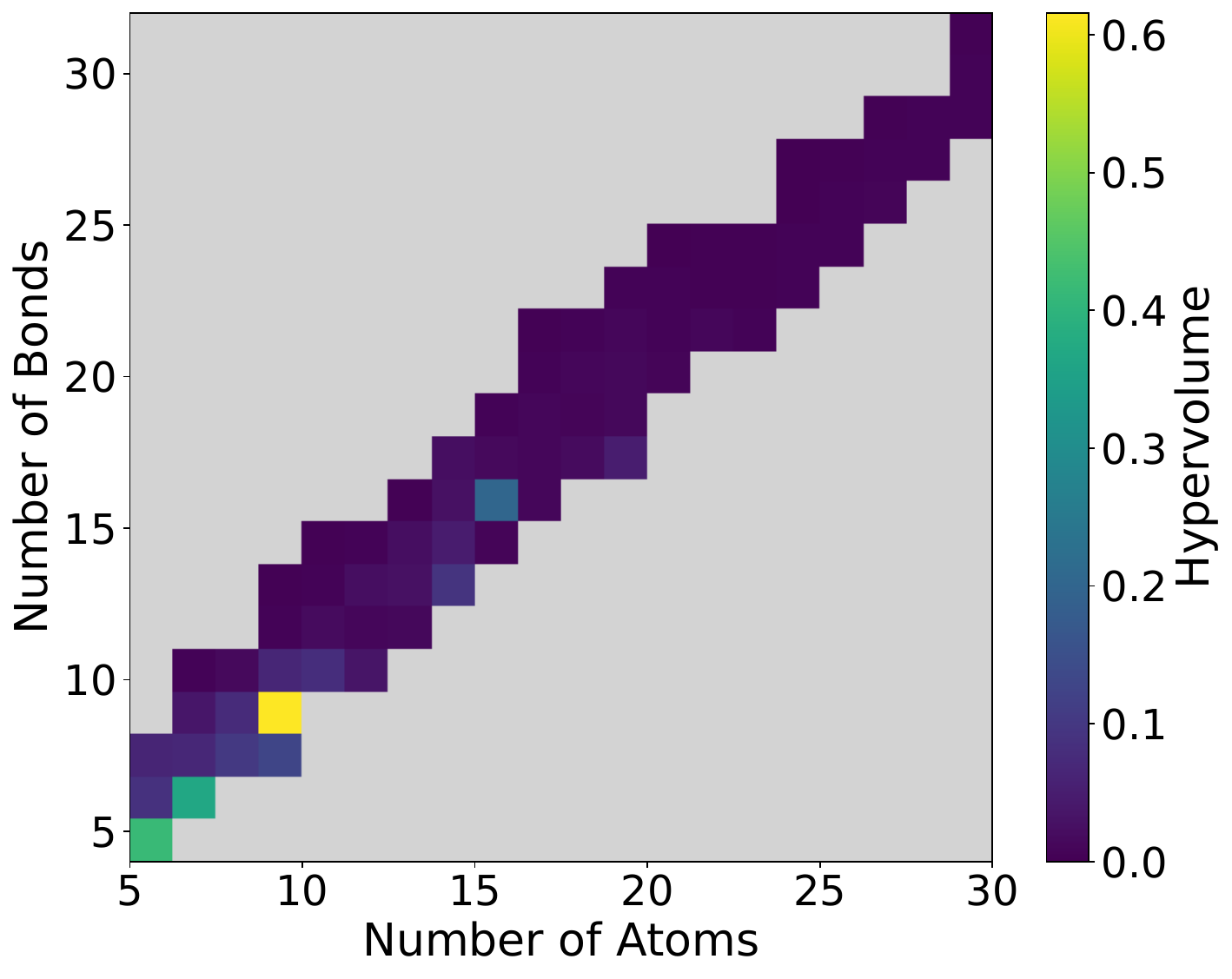}
    \caption{\small $\text{MAP-Elites}_{C}$}
    \label{fig:f_me_c_hv_heat}
\end{subfigure}\hfill
\begin{subfigure}[t]{0.14\textwidth}
    \centering
    \includegraphics[width=\linewidth]{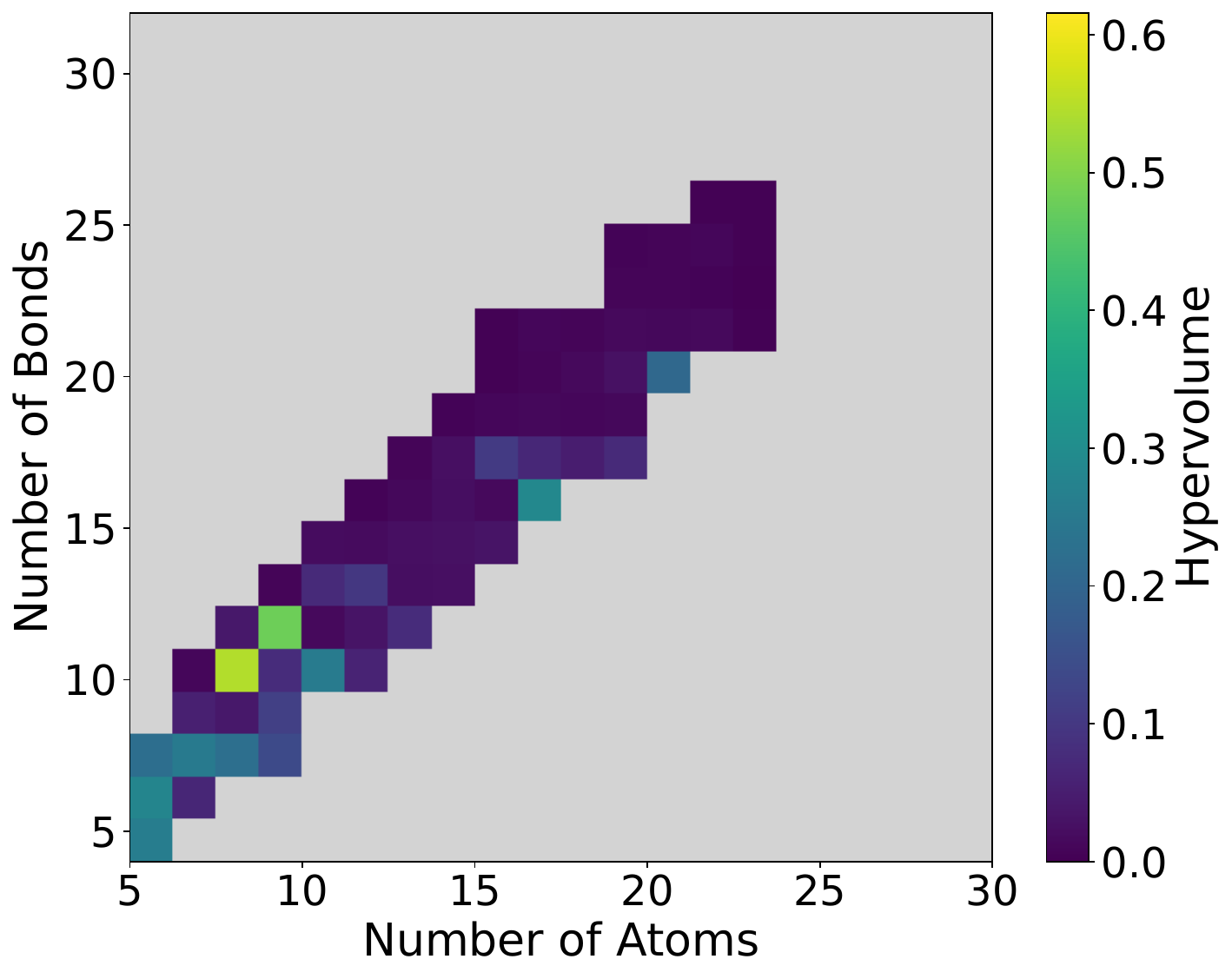}
    \caption{\small $\text{MOME}_{F}$}
    \label{fig:f_mome_f_hv_heat}
\end{subfigure}\hfill
\begin{subfigure}[t]{0.14\textwidth}
    \centering
    \includegraphics[width=\linewidth]{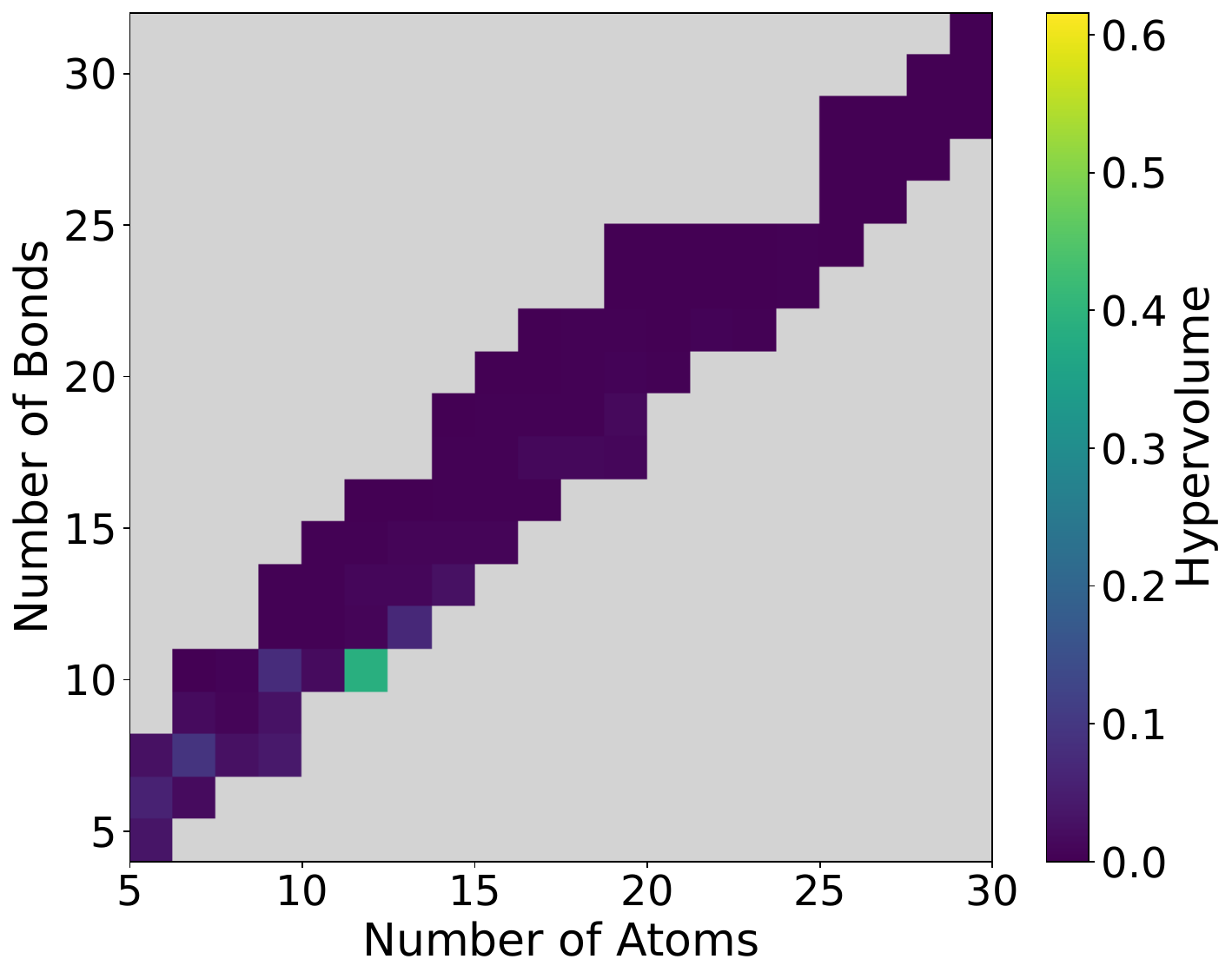}
    \caption{\small $\text{MOME}_{C}$}
    \label{fig:f_mome_c_hv_heat}
\end{subfigure}\hfill
\begin{subfigure}[t]{0.14\textwidth}
    \centering
    \includegraphics[width=\linewidth]{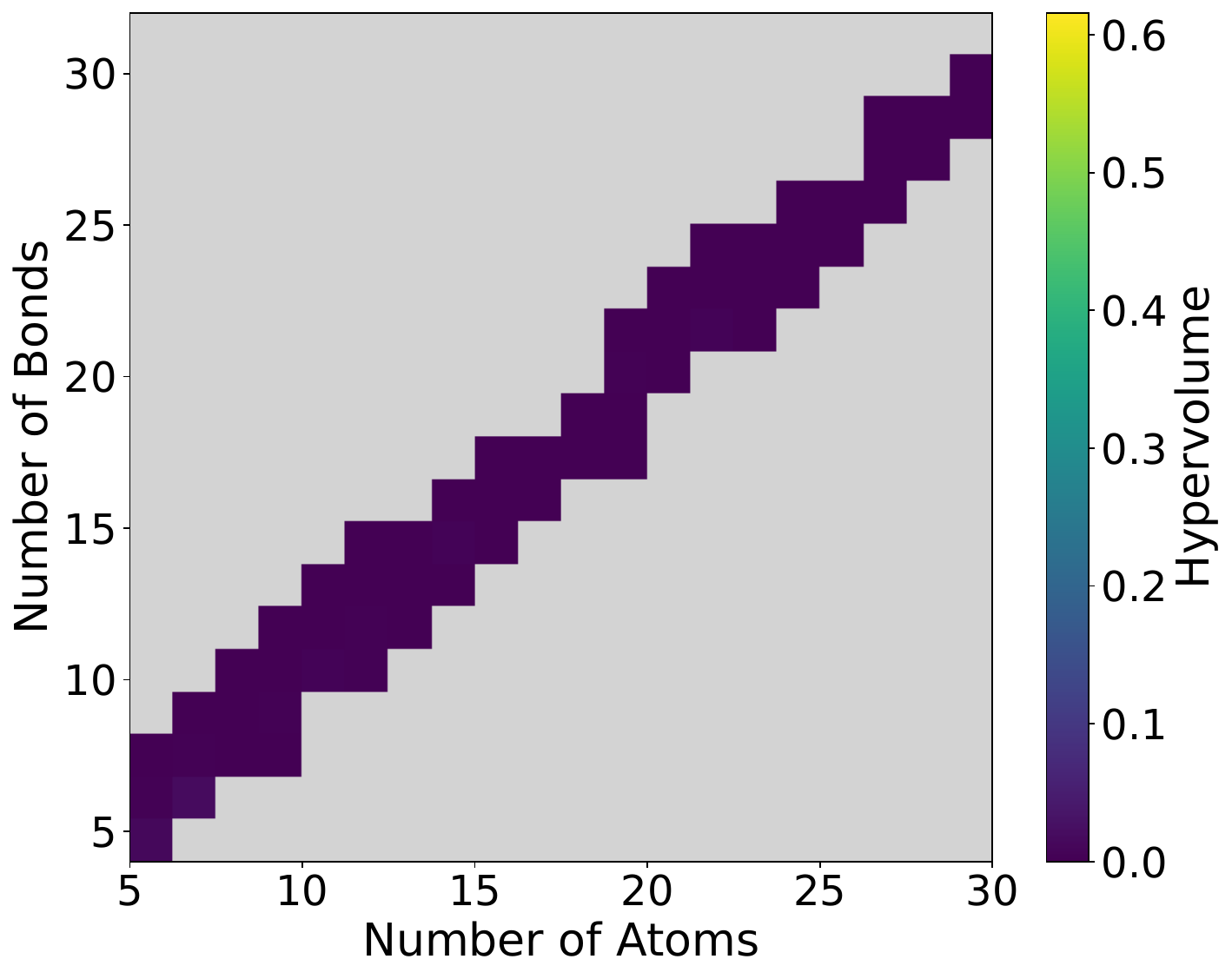}
    \caption{\small NSGA-II}
    \label{fig:f_nsga_c_hv_heat}
\end{subfigure}

\caption{\small Fine-grained Mega Archive Hypervolume Heatmaps:
Fine-grained archives that combine solutions from each algorithm across
all 20 seeds, with heat scale showing each bin's HV score. The x-axis is the atom count and the y-axis is the bond count.}
\label{fig:hv_heatmaps_f}
\Description[TODO]{TODO}
\end{figure*}
\begin{figure*}[t]
\centering
\begin{subfigure}[t]{0.14\textwidth}
    \centering
    \includegraphics[width=\linewidth]{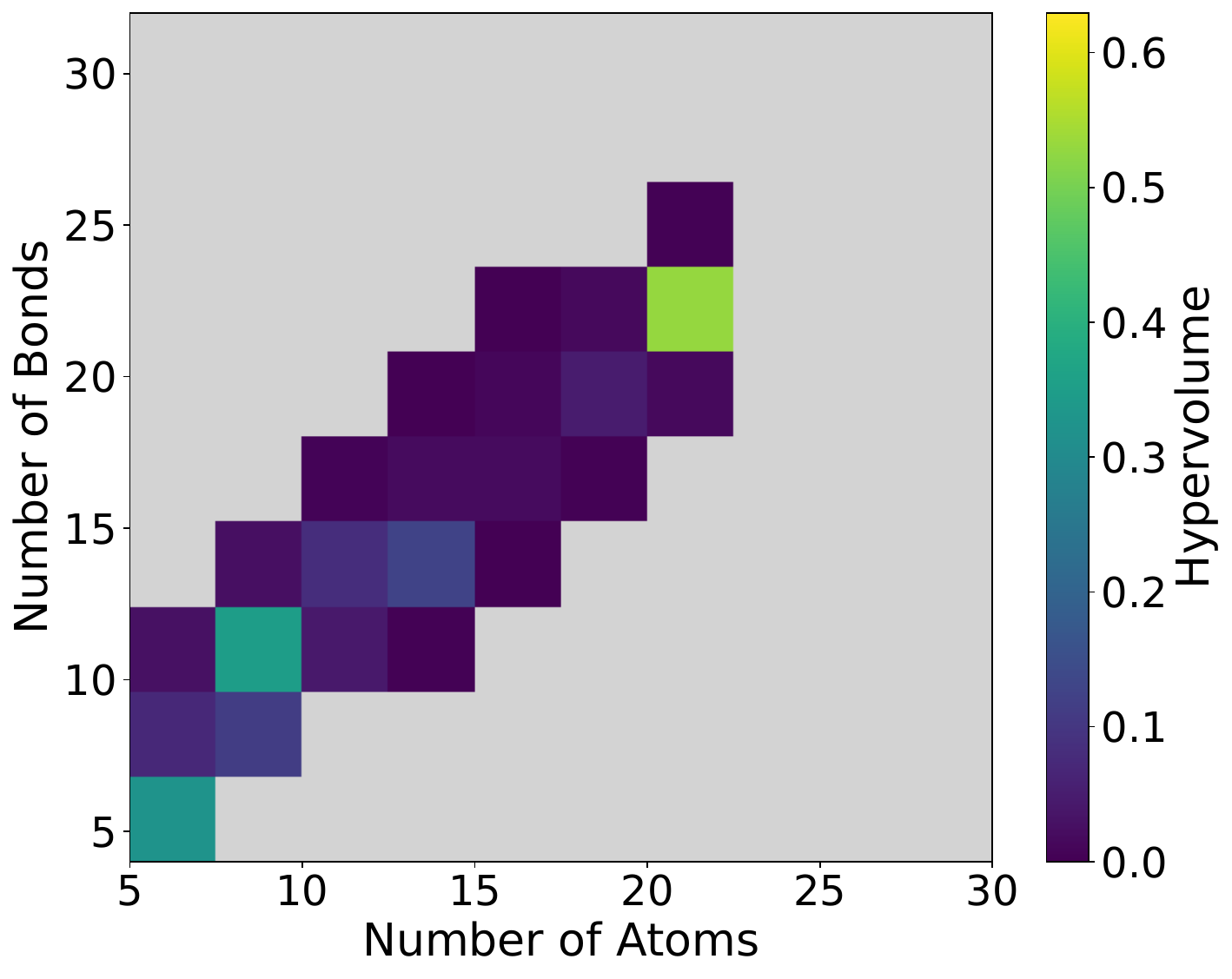}
    \caption{\small SA}
    \label{fig:c_sa_hv_heat}
\end{subfigure}\hfill
\begin{subfigure}[t]{0.14\textwidth}
    \centering
    \includegraphics[width=\linewidth]{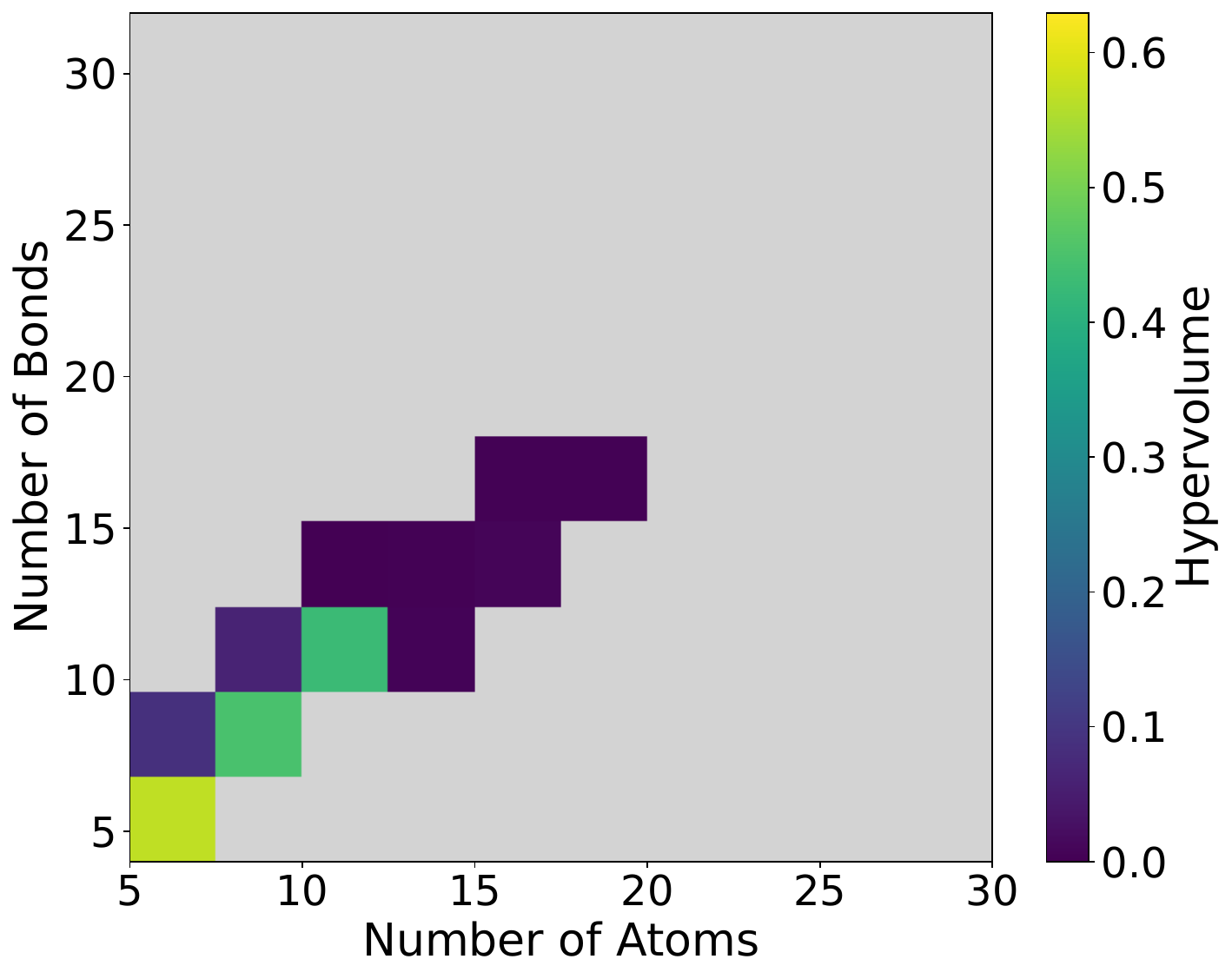}
    \caption{\small $(\mu+\lambda)$}
    \label{fig:c_ml_hv_heat}
\end{subfigure}\hfill
\begin{subfigure}[t]{0.14\textwidth}
    \centering
    \includegraphics[width=\linewidth]{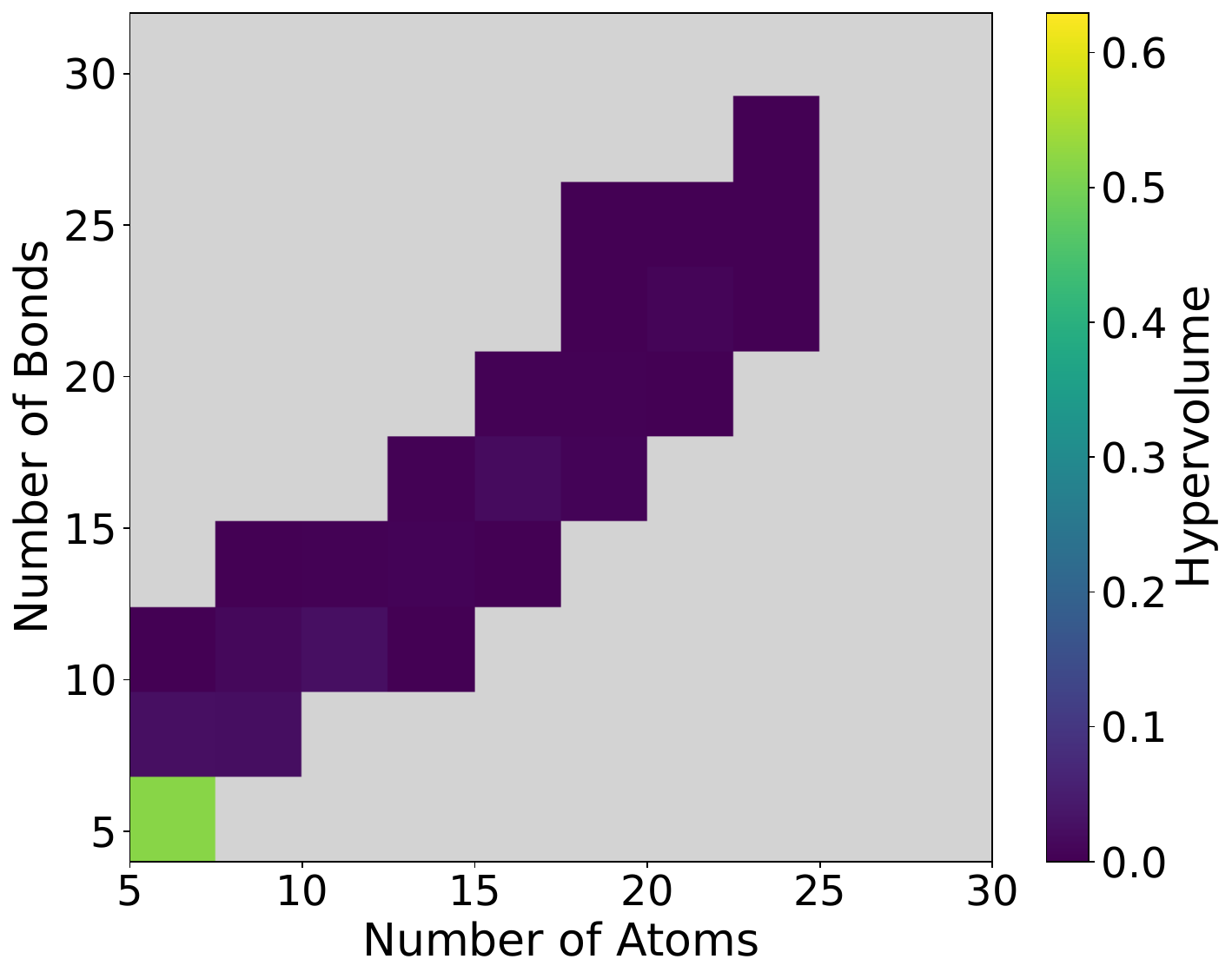}
    \caption{\small $\text{MAP-Elites}_{F}$}
    \label{fig:c_me_f_hv_heat}
\end{subfigure}\hfill
\begin{subfigure}[t]{0.14\textwidth}
    \centering
    \includegraphics[width=\linewidth]{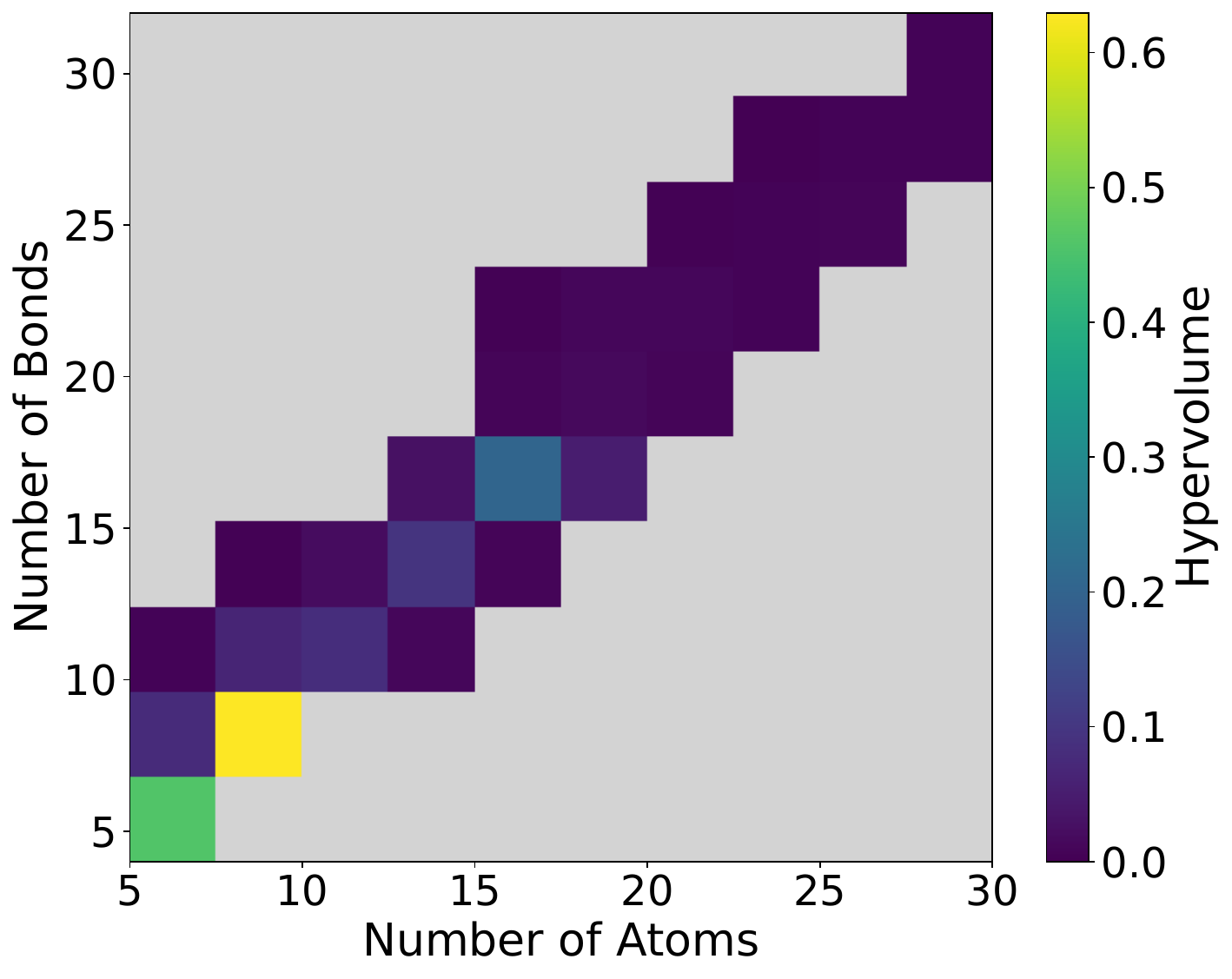}
    \caption{\small $\text{MAP-Elites}_{C}$}
    \label{fig:c_me_c_hv_heat}
\end{subfigure}\hfill
\begin{subfigure}[t]{0.14\textwidth}
    \centering
    \includegraphics[width=\linewidth]{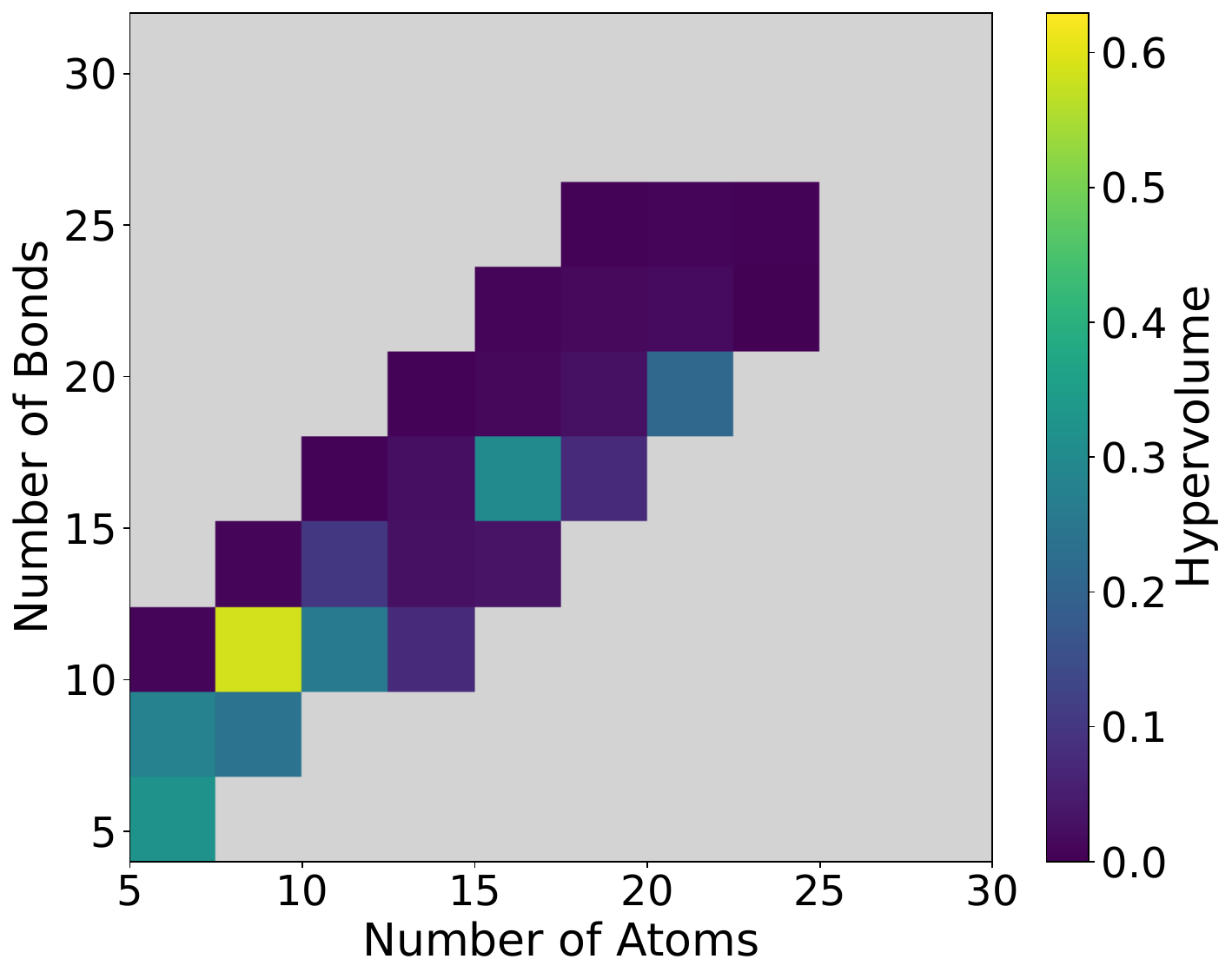}
    \caption{\small $\text{MOME}_{F}$}
    \label{fig:c_mome_f_hv_heat}
\end{subfigure}\hfill
\begin{subfigure}[t]{0.14\textwidth}
    \centering
    \includegraphics[width=\linewidth]{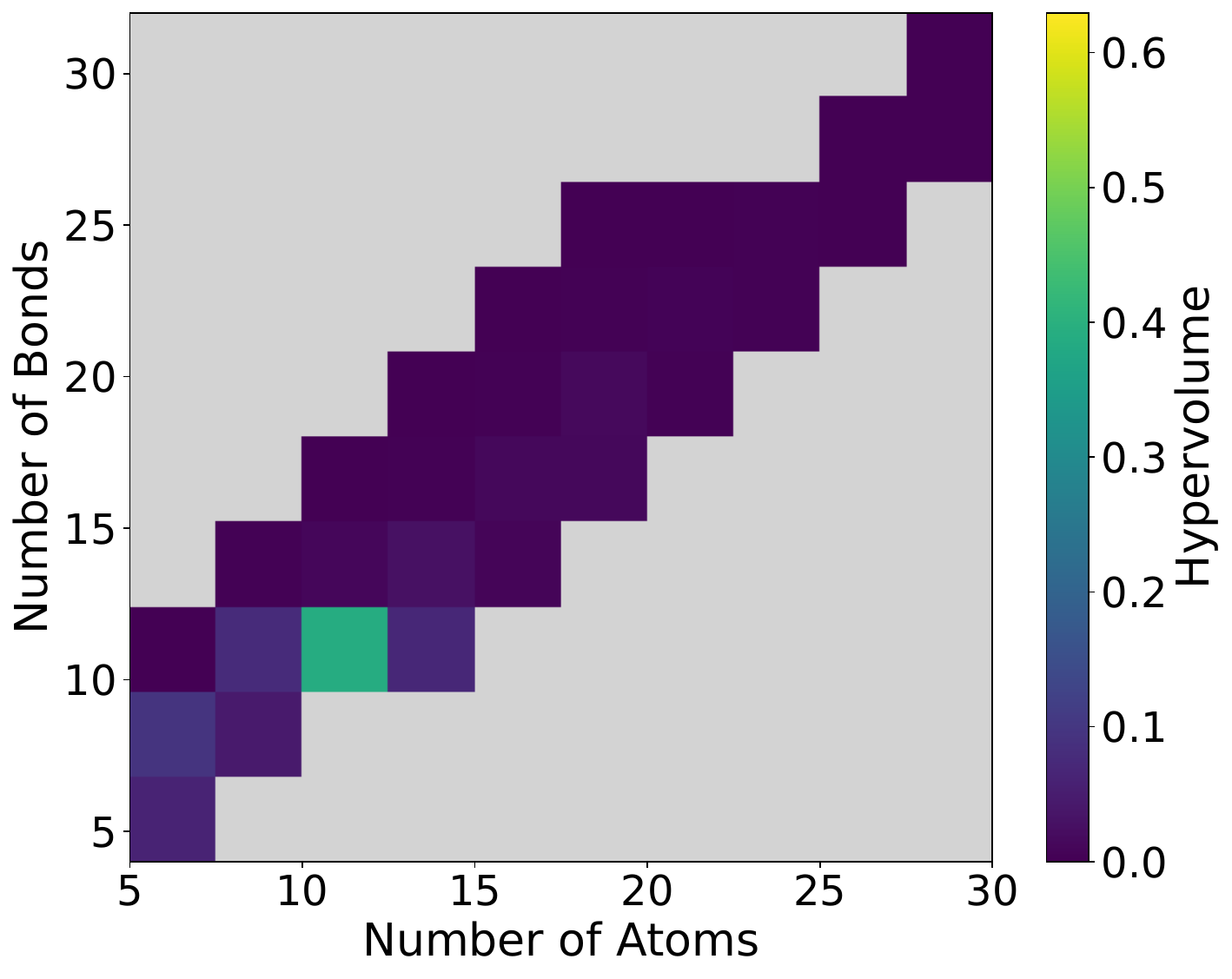}
    \caption{\small $\text{MOME}_{C}$}
    \label{fig:c_mome_c_hv_heat}
\end{subfigure}\hfill
\begin{subfigure}[t]{0.14\textwidth}
    \centering
    \includegraphics[width=\linewidth]{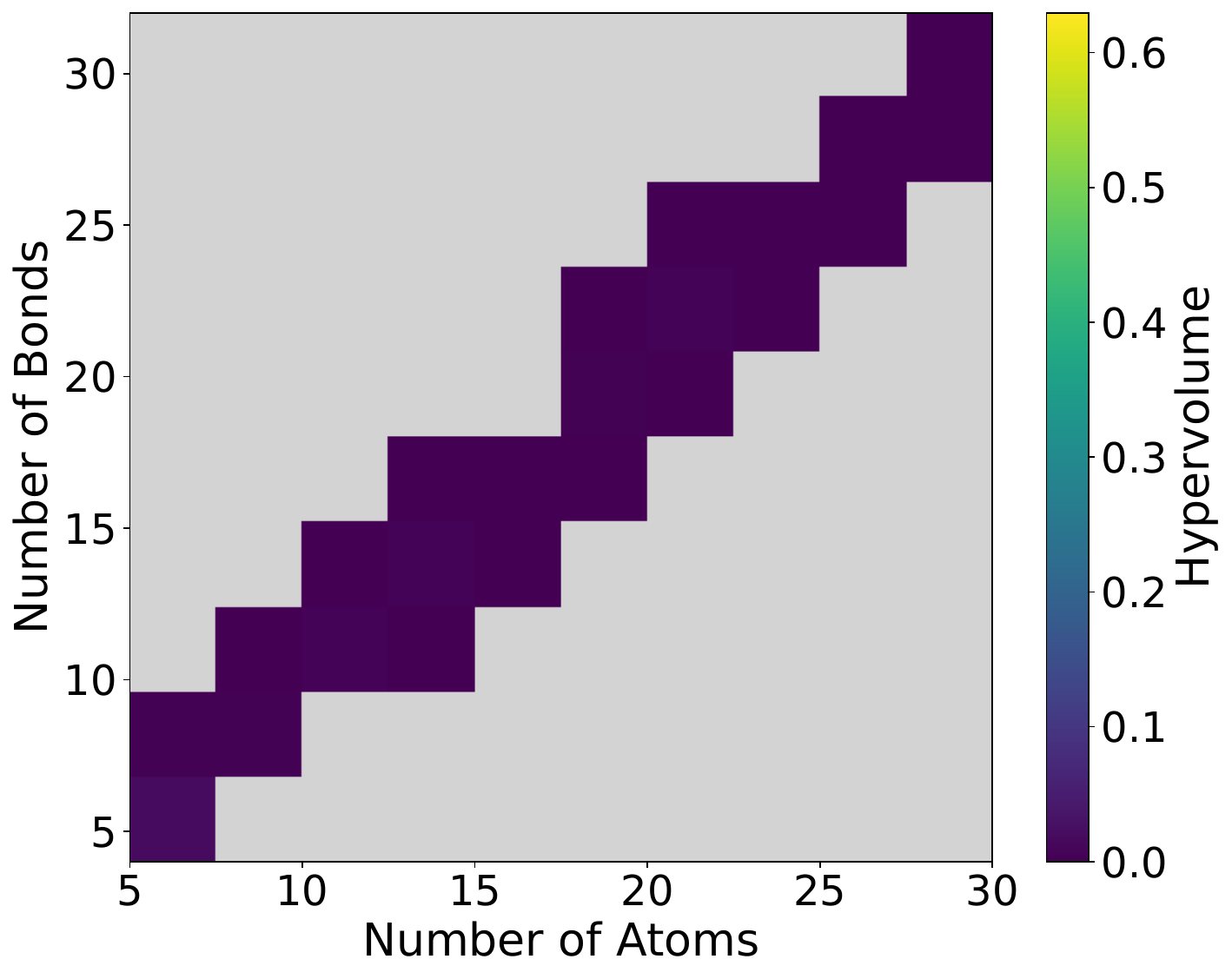}
    \caption{\small NSGA-II}
    \label{fig:c_nsga_c_hv_heat}
\end{subfigure}

\caption{\small Coarse Mega Archive Hypervolume Heatmaps:
Coarse archives that combine solutions from each algorithm across
all 20 seeds, with heat scale showing each bin's HV score. The x-axis is the atom count and the y-axis is the bond count.}
\label{fig:hv_heatmaps_c}
\Description[TODO]{TODO}
\end{figure*}

To capture overall performance, we create MOME-style archives that combine results across all 20 runs of each algorithm.
Specifically, all of the molecules generated from all runs of a given algorithm are added to both a fine (Figure \ref{fig:hv_heatmaps_f}) and coarse (Figure \ref{fig:hv_heatmaps_c}) archive, and the typical MOME rules of placement according to measure values and non-dominance apply. We then show the 
hypervolumes of individual bins in heatmaps.

All heatmaps portray a diagonal slash through the archive, because bond count and atom count are strongly correlated, but the length and width of this slash varies. 
HV scores are generally low across most bins in each archive,
with some rare bright spots.

The $(\mu+\lambda)$ archives show much less coverage, but more bright spots: these all correspond to unusually high $\beta/\gamma$ scores.
Simulated annealing has a fairly thick streak across the archive, despite low scores in many metrics above. However, the metrics above were median scores, indicating that the coverage seen in Figures~\ref{fig:f_sa_hv_heat} and~\ref{fig:c_sa_hv_heat} are the combination mostly disjoint archives from the separate runs.

$\text{MAP-Elites}_{C}$ has high MOQD and global HV scores, but the archives indicate that a large portion of this may be attributable to one very bright bin. $\text{MAP-Elites}_{F}$ has slightly less coverage, but much worse quality across most bins, explaining its lower scores.

$\text{MOME}_{F}$ has board coverage, and many bins with moderate to high HV scores, which is consistent with its high MOQD and global HV scores. It is interesting that $\text{MOME}_{F}$ has the highest coverage, global HV, and MOQD scores, even though these archives show that its coverage does not extend to the top-right of the archive as other methods do. $\text{MOME}_{C}$ extends to the top right, but performed worse in all earlier metrics. It is once again important to point out that the earlier metrics were median scores, and that the archives here combine results across all runs, thus indicating that other methods do not span the whole archive from corner to corner consistently, whereas $\text{MOME}_{F}$ does consistently cover the region shown in Figures \ref{fig:f_mome_f_hv_heat} and~\ref{fig:c_mome_f_hv_heat}

NSGA-II archives (Figures \ref{fig:f_nsga_c_hv_heat} and \ref{fig:c_nsga_c_hv_heat}) extend to the top right, but are very thin and lacking in bright spots. The lack of bright spots explains NSGA-II's low MOQD scores, and indicates that its moderately high global HV scores must depend on Pareto fronts composed of molecules from different archive bins as opposed to molecules clustered in any one bin.





\section{Discussion and Future Work}
\label{sec:discussion}

This work examines the efficacy of using single-objective, multiobjective, quality diversity, and MOQD algorithms to generate electro-optic modulators with an ideal set of properties. Our calculations show that several of these algorithms produce a diverse collection of molecules that satisfy the multiple criteria we set. In future research we plan to explore other variants of our more effective algorithms to determine how to improve the generation of molecules with specific properties.

Sometimes the quantum chemistry calculations described in Section \ref{sec:calcchem} fail, producing physically impossible results. Some of these extreme outlier results were not caught until after our experiments were completed, but we removed these extreme cases from the results analysis.
For example, the total energy per atom $E_{total}/N_{atoms}$ should not be positive, but positive values were calculated for some molecules (+668,405 Hartree in one case).
There was even one molecule with an excessively high $\beta/\gamma$ ratio over 70,000, which is impossible because it violates quantum sum rules via the Kuzyk limit~\cite{kuzyk:jstqe2001, kuzyk:physrev2000, kuzyk:optlett2000, moreno:physrev2011}.
These anomalies come from SCF convergence failures in the HF/3-21G calculations when applied to highly strained molecular geometries. Such molecules can exhibit a multi-reference character that the single-determinant Hartree-Fock theory cannot adequately describe~\cite{szabo:1996}. Rather than representing exceptional molecular properties, these values indicate limitations of the computational method, and their inclusion would artificially inflate apparent algorithm performance.

Despite achieving the highest $\beta/\gamma$ ratios among all algorithms
and more consistently high global HV scores, $(\mu+\lambda)$ selection had 
poor overall performance due to its mediocre scores in other objectives,
and low performance on Count and MOQD metrics.
Even the $\beta/\gamma$ ratios are somewhat questionable, as they reach values in the thousands which
are theoretically possible but implausible. Although we retained these results in our analysis, they are another reason why $(\mu+\lambda)$ may not be as successful as it seems.
The algorithm converges to a narrow region of chemical space with only smaller molecules that maximize the primary objective while neglecting secondary objectives. Notably, $(\mu+\lambda)$ solutions showed awful $f_{\alpha}$ compared to others that achieved perfect targeting of 0 a.u. 
The large $f_{\alpha}$ values indicate molecules that possibly exhibit increased optical absorption or dispersion, rendering them unsuitable as practical EO modulators despite strong nonlinear responses.
This demonstrates that high performance on a single metric can be misleading when the application requires molecules satisfying multiple property constraints simultaneously.

NSGA-II performed well on all objectives in isolation, and was second only to $(\mu+\lambda)$ in terms of $\beta/\gamma$ ratios. It also performed moderately well in terms of global HV, Count metrics, and MOQD scores. As a multiobjective algorithm, it produced more promising molecules exhibiting useful tradeoffs.
One example is the molecule \texttt{[O-][N+](O)OOONOCCOOO[N+]O}, which 
exhibits a linear topology dominated by oxygen-nitrogen linkages, including peroxide bridges (O-O-O) and nitro oxide functional groups. The structure achieves perfect targeting of the $f_\alpha$ constraint (0 a.u.), 
near-zero $f_{\Delta E}$ (0.80 eV),
low $E_{total}/N_{atoms}$ of -65.67 Hartree,
and a $\beta/\gamma$ ratio of 129.15, suggesting
favorable NLO response characteristics~\cite{kanis:cr1994}.

MOME's isolated median objective scores are middling, but $\text{MOME}_{F}$ has the best median global HV, Count metrics, and MOQD scores. MOME also performs well on individual QD scores for all objectives except $\beta/\gamma$ ratio, but this is an important objective. It is therefore hard to claim unequivocal success. It produces a broader range of molecules to investigate than any other method, and those molecules provide good coverage across the range of objectives.
The molecule \texttt{[C-][N+](N)=C(C)C(NN=C=N)C(=N)N=C=CON=CCC=C} is a promising example, exhibiting: $\beta/\gamma$ = 134.20, $f_\alpha$ = 14.27 a.u., $f_{\Delta E}$ = 3.87 eV, and 
$E_{total}/N_{atoms}$ = -46.10 Hartree. This molecule features an extended $\pi$-conjugated backbone incorporating multiple nitrogen heterocycles, carbodiimide groups (N=C=N), and a terminal vinyl ether moiety (ON=CCC=C). The ylide-type charge separation ([C$^-$][N$^+$]) at the molecular terminus, combined with the cumulated double bond systems, creates a push-pull electronic architecture characteristic of high-performance NLO chromophores~\cite{marder:nature1997}.
Because of the numerous trade-offs and the differing compositions of molecules, many could be useful despite not having the highest scores among all possibilities. In general, this is the promise of QD methods in general, and of MOQD methods in particular. They provide useful options for scientists to explore further.

\section{Conclusions}

We used NSGA-II, MAP-Elites, MOME, $(\mu+\lambda)$ selection, and simulated annealing to
search for SMILES strings representing molecules with desirable NLO qualities 
that could produce an effective electro-optic modulator.
Objectives included
maximization of the ratio of first-to-second hyperpolarizability $(\beta/\gamma)$, 
constraining the HOMO-LUMO gap and linear polarizability to target ranges, 
and minimization of total energy per atom.
Results show that with a fine-grained archive, MOME
covers the broadest range of structurally diverse molecules (in terms of atom and bond count),
and these molecules produce the largest global hypervolume. 
NSGA-II also produced a decent hypervolume score
and performed well on individual objectives, but scores poorly on QD/MOQD metrics.
The performance of $(\mu+\lambda)$ also seemed good, as its 
$(\beta/\gamma)$ scores were higher than all others, but this accomplishment came at the
expense of the other objectives, meaning that the molecules produced
are not useful NLO molecules, despite a high hypervolume score. We will explore the vast array of molecules produced
by all methods to assess their utility, and 
further investigate how to use these algorithms to guide the search
for molecules with desired properties.

\begin{acks}
The authors acknowledge that Generative AI (Claude) was used to refine the \texttt{matplotlib} code that produced all the result figures. 

\end{acks}

\bibliographystyle{ACM-Reference-Format}
\bibliography{mol-evo}

\clearpage

\appendix
\section{Theoretical Foundations}
\label{appendix:foundations}

This appendix provides background on the quantum chemical and computational methods used in our experiments. Sections~\ref{appendix:hf}--\ref{appendix:perturbation} cover the electronic structure theory that is the basis for our property calculations; and Section~\ref{appendix:geom} explains how molecular geometries are generated.

\subsection{The Hartree--Fock Method}
\label{appendix:hf}

The Hartree--Fock (HF) method is an \emph{ab initio} (first-principles) approach to solving the time-independent electronic Schr\"{o}dinger equation for a many-electron system. It provides the foundation for essentially all molecular orbital theory.

\subsubsection{The Electronic Problem}

Within the Born--Oppenheimer approximation, the nuclei are treated as fixed point charges and only the electronic degrees of freedom are solved quantum mechanically. The electronic Hamiltonian (in atomic units) is:
\begin{equation}
\hat{H}_{\text{elec}} = -\frac{1}{2}\sum_{i=1}^{N} \nabla_i^2 - \sum_{i=1}^{N}\sum_{A=1}^{M} \frac{Z_A}{|\vec{r}_i - \vec{R}_A|} + \sum_{i<j}^{N} \frac{1}{|\vec{r}_i - \vec{r}_j|}
\label{eq:helec}
\end{equation}
where $N$ is the number of electrons, $M$ the number of nuclei, $Z_A$ the nuclear charges, $\vec{r}_i$ the electron positions, and $\vec{R}_A$ the fixed nuclear positions. The three terms represent, respectively, the electronic kinetic energy, nuclear--electron attraction, and electron--electron repulsion.

The exact solution to $\hat{H}_{\text{elec}}\Psi = E\Psi$ is intractable for all but the smallest systems due to the electron--electron repulsion term, which couples all electrons.

\subsubsection{The Slater Determinant}

Hartree--Fock approximates the $N$-electron wavefunction as a single \emph{Slater determinant}, an anti-symmetrized product of one-electron functions (molecular orbitals, MOs) $\{\phi_i(\vec{r})\}$:
\begin{equation}
\Psi_{\text{HF}}(\vec{r}_1, \ldots, \vec{r}_N) = \frac{1}{\sqrt{N!}}
\begin{vmatrix}
\phi_1(\vec{r}_1) & \phi_2(\vec{r}_1) & \cdots & \phi_N(\vec{r}_1) \\
\phi_1(\vec{r}_2) & \phi_2(\vec{r}_2) & \cdots & \phi_N(\vec{r}_2) \\
\vdots & \vdots & \ddots & \vdots \\
\phi_1(\vec{r}_N) & \phi_2(\vec{r}_N) & \cdots & \phi_N(\vec{r}_N)
\end{vmatrix}
\label{eq:slater}
\end{equation}

The determinant form automatically satisfies the Pauli exclusion principle: exchanging any two electrons changes the sign of $\Psi$. Each electron observes the average field of the other $N-1$ electrons rather than their instantaneous positions; this is the \emph{mean-field approximation}, and the difference between the HF energy and the exact energy is called the \emph{correlation energy}.

\subsubsection{The Roothaan--Hall Equations}

Each MO $\phi_i$ is expanded in a set of known basis functions $\{\chi_\mu\}$ (see Section~\ref{appendix:basis}):
\begin{equation}
\phi_i(\vec{r}) = \sum_{\mu=1}^{K} C_{\mu i}\, \chi_\mu(\vec{r})
\label{eq:lcao}
\end{equation}
where $K$ is the number of basis functions and $C_{\mu i}$ are the expansion coefficients to be determined. Substituting into the variational condition $\delta E[\Psi_{\text{HF}}] = 0$ yields the Roothaan--Hall matrix eigenvalue equation:
\begin{equation}
\mathbf{F}\mathbf{C} = \mathbf{S}\mathbf{C}\boldsymbol{\epsilon}
\label{eq:roothaan}
\end{equation}
where:
\begin{itemize}
\item $\mathbf{F}$ is the \emph{Fock matrix}, with elements: \\ $F_{\mu\nu} = H_{\mu\nu}^{\text{core}} + \sum_{\lambda\sigma} D_{\lambda\sigma}\left[(\mu\nu|\lambda\sigma) - \tfrac{1}{2}(\mu\lambda|\nu\sigma)\right]$
\item $H_{\mu\nu}^{\text{core}} = \langle\chi_\mu|-\tfrac{1}{2}\nabla^2 - \sum_A Z_A/|\vec{r}-\vec{R}_A||\chi_\nu\rangle$ is the one-electron (core) Hamiltonian
\item $D_{\lambda\sigma} = 2\sum_{i=1}^{N/2} C_{\lambda i}C_{\sigma i}$ is the density matrix (for closed-shell systems)
\item $(\mu\nu|\lambda\sigma) = \iint \chi_\mu(\vec{r}_1)\chi_\nu(\vec{r}_1)\frac{1}{r_{12}}\chi_\lambda(\vec{r}_2)\chi_\sigma(\vec{r}_2)\,d\vec{r}_1\,d\vec{r}_2$ are two-electron integrals
\item $\mathbf{S}$ is the overlap matrix, $S_{\mu\nu} = \langle\chi_\mu|\chi_\nu\rangle$
\item $\mathbf{C}$ is the matrix of MO coefficients
\item $\boldsymbol{\epsilon}$ is the diagonal matrix of orbital energies
\end{itemize}

\subsubsection{The Self-Consistent Field (SCF) Procedure}

Equation~\ref{eq:roothaan} is nonlinear because $\mathbf{F}$ depends on $\mathbf{D}$, which depends on $\mathbf{C}$, which is the solution. It is solved iteratively:

\begin{enumerate}
\item Guess an initial density matrix $\mathbf{D}^{(0)}$ (e.g., from a core Hamiltonian diagonalization).
\item Build the Fock matrix $\mathbf{F}^{(n)}$ from $\mathbf{D}^{(n)}$.
\item Solve $\mathbf{F}^{(n)}\mathbf{C}^{(n)} = \mathbf{S}\mathbf{C}^{(n)}\boldsymbol{\epsilon}^{(n)}$ for new orbitals.
\item Construct a new density matrix $\mathbf{D}^{(n+1)}$ from the occupied orbitals in $\mathbf{C}^{(n)}$.
\item Check convergence: if $|E^{(n+1)} - E^{(n)}| < \tau$ (we use $\tau = 10^{-9}$~Ha), stop. Otherwise, return to step~2.
\end{enumerate}

Upon convergence, the total HF energy is:
\begin{equation}
E_{\text{HF}} = \frac{1}{2}\sum_{\mu\nu} D_{\mu\nu}(H_{\mu\nu}^{\text{core}} + F_{\mu\nu}) + E_{\text{nn}}
\end{equation}
where $E_{\text{nn}} = \sum_{A<B} Z_A Z_B / |\vec{R}_A - \vec{R}_B|$ is the nuclear--nuclear repulsion energy.

\subsubsection{Restricted vs.\ Unrestricted HF}

For closed-shell molecules (all electrons paired), we use \emph{restricted} HF (RHF), where each spatial orbital is doubly occupied by one $\alpha$-spin and one $\beta$-spin electron. For open-shell molecules (unpaired electrons), we use \emph{unrestricted} HF (UHF), which allows separate spatial orbitals for $\alpha$ and $\beta$ electrons, resulting in two coupled Fock equations.

\subsection{Basis Sets}
\label{appendix:basis}

The basis functions $\{\chi_\mu\}$ in Equation~\ref{eq:lcao} determine the flexibility of the MO expansion and thus the accuracy of the calculation. We use the 3-21G basis set~\cite{binkley1980self}, a \emph{split-valence} basis of Gaussian-type orbitals (GTOs).

\subsubsection{Gaussian-Type Orbitals}

A Cartesian GTO centered on nucleus $A$ has the form:
\begin{equation}
\chi(\vec{r}) = N\, x_A^l\, y_A^m\, z_A^n\, e^{-\zeta |\vec{r} - \vec{R}_A|^2}
\end{equation}
where $x_A = x - R_{Ax}$, etc., $N$ is a normalization constant, $l + m + n$ determines the angular momentum ($s$-type for $l+m+n=0$, $p$-type for $l+m+n=1$, etc.), and $\zeta$ is the orbital exponent controlling radial extent. GTOs are computationally advantageous because the product of two Gaussians centered at different points is another Gaussian, enabling efficient evaluation of the two-electron integrals.

\subsubsection{Contracted GTOs and the 3-21G Basis}

Rather than using single GTOs, basis sets use fixed linear combinations called \emph{contracted} GTOs. The notation ``3-21G'' encodes the contraction scheme:
\begin{itemize}
\item Core orbitals: Each inner-shell orbital is represented by a single contracted function composed of 3 primitive Gaussians.
\item Valence orbitals: Each valence shell is divided into two contracted functions, an inner part composed of 2 primitives and an outer part composed of 1 primitive.
\end{itemize}

This split-valence scheme allows the valence region (where chemical bonding occurs) to be described more flexibly than the core, at minimal additional cost. For example, a carbon atom in 3-21G has:
\begin{itemize}
\item 1$s$ core: 1 contracted function (3 primitives)
\item 2$s$, 2$p$ valence: 2 contracted functions each (2+1 primitives), totaling 8 valence basis functions ($2s + 6p$)
\end{itemize}
giving 9 basis functions per carbon atom.

\subsubsection{Selection of 3-21G}

The 3-21G basis is a minimal split-valence set---small enough to evaluate thousands of molecules (each requiring ${\sim}30$ SCF calculations for the finite-field procedure), yet accurate enough for reliable \emph{relative} rankings of molecular hyperpolarizabilities. Systematic benchmarking~\cite{mashak:arxiv2025} demonstrated 100\% pairwise ranking consistency against experimental $\beta$ values for a test set of molecules, even though absolute values differ from experiment. Since evolutionary algorithms rely on ranking (via tournament selection, elitist replacement, or Pareto dominance) rather than absolute values, preserving the correct ordering is sufficient for evolutionary purposes.

\subsection{Perturbation Theory and the Definition of Optical Response Properties}
\label{appendix:perturbation}

The polarizability ($\alpha$), first hyperpolarizability ($\beta$), and second hyperpolarizability ($\gamma$) arise naturally from expanding the molecular energy and dipole moment as Taylor series in the applied electric field.

\subsubsection{Energy Expansion}

When a static uniform electric field $\vec{E}$ is applied to a molecule, the total energy can be expanded as:
\begin{equation}
\begin{split}
E(\vec{E}) = E^{(0)} - \sum_i \mu_i^{(0)} E_i - \frac{1}{2}\sum_{ij} \alpha_{ij} E_i E_j \\- \frac{1}{6}\sum_{ijk} \beta_{ijk} E_i E_j E_k - \frac{1}{24}\sum_{ijkl} \gamma_{ijkl} E_i E_j E_k E_l - \cdots
\label{eq:energy_expansion}
\end{split}
\end{equation}
where $E^{(0)}$ is the field-free energy, $\mu_i^{(0)}$ is the permanent dipole moment, and the sums run over Cartesian indices $\{x, y, z\}$. The coefficients $\alpha_{ij}$, $\beta_{ijk}$, and $\gamma_{ijkl}$ are, respectively, the linear polarizability, first hyperpolarizability, and second hyperpolarizability tensors.

\subsubsection{Dipole Moment Expansion}

Equivalently, the field-dependent dipole moment is:
\begin{equation}
\mu_i(\vec{E}) = \mu_i^{(0)} + \sum_j \alpha_{ij} E_j + \frac{1}{2}\sum_{jk} \beta_{ijk} E_j E_k + \frac{1}{6}\sum_{jkl} \gamma_{ijkl} E_j E_k E_l + \cdots
\label{eq:dipole_expansion}
\end{equation}

since $\mu_i = -\partial E / \partial E_i$. This relationship is the basis for the finite-field method: by computing $\mu_i(\vec{E})$ at several field values and applying finite difference formulas, we extract $\alpha$, $\beta$, and $\gamma$ numerically without needing to solve the perturbation theory equations analytically.

\subsubsection{Physical Interpretation}

\begin{itemize}
\item $\alpha_{ij}$ (linear polarizability): How easily an electric field distorts the electron cloud. Molecules with delocalized $\pi$-electrons (e.g., conjugated systems) tend to have large $\alpha$. Units: $e^2 a_0^2 / E_h$ (a.u.), or equivalently volume in \AA$^3$ ($1$~a.u.~$= 0.1482$~\AA$^3$).

\item $\beta_{ijk}$ (first hyperpolarizability): Second-order NLO effects such as second-harmonic generation (frequency doubling) and the Pockels effect (linear electro-optic effect). $\beta$ is nonzero only in molecules lacking an inversion center (noncentrosymmetric molecules), which is why our search targets asymmetric organic chromophores. Units: $e^3 a_0^3 / E_h^2$ (a.u.).

\item $\gamma_{ijkl}$ (second hyperpolarizability): Third-order effects such as the optical Kerr effect, self-phase modulation, and third-harmonic generation. Unlike $\beta$, $\gamma$ is nonzero for all molecules. Units: $e^4 a_0^4 / E_h^3$ (a.u.).
\end{itemize}

The $\beta/\gamma$ ratio thus quantifies the relative strength of second-order vs.\ third-order responses. A high ratio indicates a molecule whose NLO behavior is dominated by the desirable Pockels effect rather than unintended third-order effects.

\subsubsection{Connection to the Finite-Field Method}

From Equation~\ref{eq:dipole_expansion}, applying a field $E_j = h$ along axis $j$ and differentiating numerically yields $\alpha$ (first derivative), $\beta$ (second derivative), and from the energy expansion (Equation~\ref{eq:energy_expansion}), $\gamma$ (fourth derivative). The specific finite difference formulas used are given in Sections~\ref{appendix:alpha_calc}--\ref{appendix:gamma_calc}.

\subsection{Canonical SMILES}

A given molecule can be represented by multiple valid SMILES strings (e.g., \texttt{OCC} and \texttt{CCO} both encode ethanol). To ensure unique representations for comparison and deduplication, we use \emph{canonical} SMILES generated by RDKit~\cite{landrum:rdkit2010}, which applies a deterministic algorithm to produce exactly one SMILES string per molecular graph.

\subsection{Molecular Geometry Generation}
\label{appendix:geom}

Quantum chemical calculations require 3D atomic coordinates, but SMILES strings encode only the molecular graph (connectivity). The conversion proceeds in three steps using RDKit~\cite{landrum:rdkit2010}:

\begin{enumerate}
\item 2D coordinate generation: RDKit assigns approximate 2D coordinates using a template-based algorithm.
\item 3D embedding: The \texttt{EmbedMolecule} function generates initial 3D coordinates using distance geometry~\cite{blaney:armc1991}, which samples conformations consistent with bond lengths, bond angles, and torsional preferences derived from empirical rules.
\item Force-field optimization: The 3D structure is refined using the Universal Force Field (UFF)~\cite{rappe:jacs1992} or MMFF94~\cite{halgren:jcc1996}, which minimizes the classical strain energy to produce a reasonable starting geometry.
\end{enumerate}

The resulting geometry is \emph{not} a quantum-mechanically optimized structure; it is a classical mechanics approximation. Full geometry optimization at the HF/3-21G level would require gradient calculations at each optimization step, multiplying the computational cost by a factor of ${\sim}10$--$50$ per molecule, which is prohibitive when evaluating thousands of molecules. Benchmarking~\cite{mashak:arxiv2025} showed that RDKit geometries preserve the relative ranking of $\beta$ values, which is sufficient for evolutionary selection.


\section{Property Calculations}
\label{appendix:calculations}

This section provides the complete mathematical specification of each property calculation. All properties are computed at the HF/3-21G level using PySCF~\cite{sun:jcphys2020}, as described in Appendix~\ref{appendix:hf}.

\subsection{Electric Field Perturbation}
\label{appendix:field}

The finite field method computes NLO response properties by numerical differentiation of observables (dipole moments or total energies) with respect to an applied static electric field. The external field $\vec{E} = (E_x, E_y, E_z)$ is incorporated into the one-electron Hamiltonian as
\begin{equation}
\hat{H}_{\text{core}}' = \hat{H}_{\text{core}} + \sum_{\alpha=x,y,z} E_\alpha \hat{r}_\alpha
\label{eq:hcore_field}
\end{equation}
where $\hat{r}_\alpha$ is the $\alpha$-component of the position operator. In matrix form, the modified core Hamiltonian in the atomic orbital basis is
\begin{equation}
H'_{\mu\nu} = H^{(0)}_{\mu\nu} + \sum_{\alpha} E_\alpha \langle \mu | \hat{r}_\alpha | \nu \rangle
\end{equation}
where $H^{(0)}_{\mu\nu}$ is the field-free core Hamiltonian (kinetic energy plus nuclear attraction) and $\langle \mu | \hat{r}_\alpha | \nu \rangle$ are the electric dipole integrals. The self-consistent field (SCF) procedure then converges to a new set of molecular orbitals and density matrix in the presence of the field. A tight convergence threshold of $10^{-9}$~a.u.\ on the total energy is used to ensure that the finite differences are numerically stable.

All finite field calculations use a field strength of $h = 0.001$~a.u.\ ($\approx 5.14 \times 10^{6}$~V/m), which balances numerical precision against higher-order contamination~\cite{kurtz:jcc1990}.

\subsection{Dipole Moment}
\label{appendix:dipole}

The total dipole moment $\vec{\mu}$ is the sum of electronic and nuclear contributions:
\begin{equation}
\mu_\alpha = \mu_\alpha^{\text{elec}} + \mu_\alpha^{\text{nuc}}
\end{equation}
where the electronic contribution is computed from the one-particle density matrix $D_{\mu\nu}$ and the dipole integrals:
\begin{equation}
\mu_\alpha^{\text{elec}} = -\sum_{\mu\nu} D_{\mu\nu} \langle \mu | \hat{r}_\alpha | \nu \rangle = -\text{Tr}[\mathbf{D} \cdot \mathbf{r}_\alpha]
\end{equation}
and the nuclear contribution is
\begin{equation}
\mu_\alpha^{\text{nuc}} = \sum_{A} Z_A R_{A,\alpha}
\end{equation}
with $Z_A$ and $R_{A,\alpha}$ being the charge and $\alpha$-coordinate of nucleus $A$.

\subsection{Linear Polarizability ($\alpha$)}
\label{appendix:alpha_calc}

The static linear polarizability tensor $\alpha_{ij}$ describes the first-order change in dipole moment with respect to an applied electric field:
\begin{equation}
\alpha_{ij} = \frac{\partial \mu_i}{\partial E_j}
\end{equation}

We compute $\alpha_{ij}$ numerically via central finite differences on the dipole moment. For each Cartesian axis $j \in \{x, y, z\}$, a field of magnitude $\pm h$ is applied along axis $j$, yielding:
\begin{equation}
\alpha_{ij} \approx \frac{\mu_i(+h\hat{e}_j) - \mu_i(-h\hat{e}_j)}{2h}
\label{eq:alpha_fd}
\end{equation}
where $\mu_i(+h\hat{e}_j)$ denotes the $i$-th component of the dipole moment computed with an electric field of $+h$ applied along the $j$-th axis. This requires 7 single-point calculations: one at zero field and one at $\pm h$ along each of the three Cartesian axes.

The full $3\times 3$ polarizability tensor is thus:
\begin{equation}
\boldsymbol{\alpha} = \begin{pmatrix}
\alpha_{xx} & \alpha_{xy} & \alpha_{xz} \\
\alpha_{yx} & \alpha_{yy} & \alpha_{yz} \\
\alpha_{zx} & \alpha_{zy} & \alpha_{zz}
\end{pmatrix}
\end{equation}

The isotropic (rotationally averaged) mean polarizability used in the objective function is:
\begin{equation}
\bar{\alpha} = \frac{1}{3}\text{Tr}(\boldsymbol{\alpha}) = \frac{\alpha_{xx} + \alpha_{yy} + \alpha_{zz}}{3}
\label{eq:alpha_mean}
\end{equation}

\subsection{First Hyperpolarizability ($\beta$)}
\label{appendix:beta_calc}

The static first hyperpolarizability tensor $\beta_{ijk}$ is the second-order response of the dipole moment to an applied electric field:
\begin{equation}
\beta_{ijk} = \frac{\partial^2 \mu_i}{\partial E_j \partial E_k}
\end{equation}

We compute the full $3\times 3\times 3$ tensor (all 27 elements $\beta_{ijk}$) using second-order finite differences on the dipole moment~\cite{kurtz:jcc1990, bishop:jws1998}.

\subsubsection{Diagonal Elements ($\beta_{iii}$)}

The three diagonal elements ($i = j = k$) are computed using the standard three-point central difference formula:
\begin{equation}
\beta_{iii} = -\frac{\mu_i(+h\hat{e}_i) - 2\mu_i(\vec{0}) + \mu_i(-h\hat{e}_i)}{h^2}
\label{eq:beta_diag}
\end{equation}
where $\mu_i(\vec{0})$ is the $i$-th dipole component at zero field:
\begin{align}
\beta_{xxx} &= -\frac{\mu_x(+h,0,0) - 2\mu_x(0,0,0) + \mu_x(-h,0,0)}{h^2} \\
\beta_{yyy} &= -\frac{\mu_y(0,+h,0) - 2\mu_y(0,0,0) + \mu_y(0,-h,0)}{h^2} \\
\beta_{zzz} &= -\frac{\mu_z(0,0,+h) - 2\mu_z(0,0,0) + \mu_z(0,0,-h)}{h^2}
\end{align}

\subsubsection{Off-Diagonal Elements with Two Equal Indices ($\beta_{iij}$)}

Elements with exactly two equal indices are computed using a four-point mixed finite difference formula:
\begin{align}
    &\beta_{iij} =\\
    &-\frac{\mu_i(+h\hat{e}_i + h\hat{e}_j) - \mu_i(+h\hat{e}_i - h\hat{e}_j) - \mu_i(-h\hat{e}_i + h\hat{e}_j) + \mu_i(-h\hat{e}_i - h\hat{e}_j)}{4h^2}
\label{eq:beta_offdiag}
\end{align}
for $i \neq j$. Due to Kleinman symmetry~\cite{kanis:cr1994} (permutation symmetry of the static tensor), all permutations of equal indices are assigned the same value:
\begin{equation}
\beta_{iij} = \beta_{iji} = \beta_{jii}
\end{equation}

For example:
\begin{align}
&\beta_{xxy} =\\ 
&-\frac{\mu_x(+h,+h,0) - \mu_x(+h,-h,0) - \mu_x(-h,+h,0) + \mu_x(-h,-h,0)}{4h^2}
\end{align}
and $\beta_{xyx} = \beta_{yxx} = \beta_{xxy}$.

\subsubsection{Fully Off-Diagonal Element ($\beta_{xyz}$)}

The element with all three indices distinct is computed as:
\begin{align}
&\beta_{xyz} =\\
&-\frac{\mu_x(0,+h,+h) - \mu_x(0,+h,-h) - \mu_x(0,-h,+h) + \mu_x(0,-h,-h)}{4h^2}
\label{eq:beta_xyz}
\end{align}
By Kleinman symmetry, all six permutations of $(x,y,z)$ share this value:
\begin{equation}
\beta_{xyz} = \beta_{xzy} = \beta_{yxz} = \beta_{yzx} = \beta_{zxy} = \beta_{zyx}
\end{equation}

\subsubsection{Total Number of Field Configurations}

The full tensor calculation requires the following electric field configurations:
\begin{itemize}
\item 1 zero-field calculation: $(0, 0, 0)$
\item 6 single-axis calculations: $(\pm h, 0, 0)$, $(0, \pm h, 0)$, $(0, 0, \pm h)$
\item 12 two-axis calculations: $(\pm h, \pm h, 0)$, $(\pm h, 0, \pm h)$, $(0, \pm h, \pm h)$
\end{itemize}
This gives 19 distinct field configurations for the $\beta$ tensor. An additional 6 single-point calculations at $\pm 2h$ along each axis are needed for the $\gamma$ calculation (Section~\ref{appendix:gamma_calc}), for a total of 25 distinct field configurations per molecule. Some additional configurations are computed due to permutation bookkeeping, yielding approximately 27 SCF calculations per molecule for $\beta$ alone.

\subsubsection{Derived Scalar Quantities}

From the full tensor, we compute two scalar measures of the hyperpolarizability.

\paragraph{Vector Hyperpolarizability ($\beta_{\text{vec}}$).} The vector component along each Cartesian direction is defined as~\cite{kanis:cr1994}:
\begin{equation}
\beta_i = \frac{1}{3}\left(\beta_{iii} + \beta_{ijj} + \beta_{ikk}\right), \quad i \neq j \neq k
\label{eq:beta_vector_comp}
\end{equation}
Explicitly:
\begin{align}
\beta_x &= \frac{1}{3}(\beta_{xxx} + \beta_{xyy} + \beta_{xzz}) \\
\beta_y &= \frac{1}{3}(\beta_{yyy} + \beta_{yxx} + \beta_{yzz}) \\
\beta_z &= \frac{1}{3}(\beta_{zzz} + \beta_{zxx} + \beta_{zyy})
\end{align}
The vector magnitude is:
\begin{equation}
\beta_{\text{vec}} = \sqrt{\beta_x^2 + \beta_y^2 + \beta_z^2}
\label{eq:beta_vec}
\end{equation}

\paragraph{Mean Hyperpolarizability ($\beta_{\text{mean}}$).} The rotationally averaged (isotropic) mean hyperpolarizability is~\cite{bishop:jws1998, kanis:cr1994}:
\begin{equation}
\beta_{\text{mean}} = \frac{1}{3}\left[\sum_{i} \beta_{iii} + 2\sum_{\substack{i,j \\ i \neq j}} \beta_{ijj}\right]
\label{eq:beta_mean}
\end{equation}
Expanding:
\begin{equation}
\begin{split}
&\beta_{\text{mean}} =\frac{1}{3}\Big[\beta_{xxx} + \beta_{yyy} + \beta_{zzz} +\\
&2(\beta_{xyy} + \beta_{xzz} + \beta_{yxx} + \beta_{yzz} + \beta_{zxx} + \beta_{zyy})\Big]
\end{split}
\end{equation}

This is the value of $\beta$ used in the $\beta/\gamma$ objective. Negative values of $\beta_{\text{mean}}$ (which can arise from numerical noise in near-zero cases) are clamped to zero.

\subsection{Second Hyperpolarizability ($\gamma$)}
\label{appendix:gamma_calc}

The static second hyperpolarizability tensor $\gamma_{ijkl}$ is the fourth-order response of the total energy to an applied electric field:
\begin{equation}
\gamma_{ijkl} = -\frac{\partial^4 E_{\text{tot}}}{\partial E_i \partial E_j \partial E_k \partial E_l}
\end{equation}

We compute the diagonal components $\gamma_{iiii}$ ($i \in \{x,y,z\}$) using a five-point central difference formula on the total energy~\cite{kurtz:jcc1990}:
\begin{equation}
\gamma_{iiii} = -\frac{E(-2h\hat{e}_i) - 4E(-h\hat{e}_i) + 6E(\vec{0}) - 4E(+h\hat{e}_i) + E(+2h\hat{e}_i)}{h^4}
\label{eq:gamma_fd}
\end{equation}
where $E(\vec{F})$ is the SCF total energy computed with external field $\vec{F}$. This is the standard five-point stencil approximation for the fourth derivative. The required field strengths are $0, \pm h, \pm 2h$ along each axis, where the zero-field and $\pm h$ energies are reused from the $\beta$ calculation.

Explicitly:
\begin{align}
&\gamma_{xxxx} =\\ 
&-\frac{E(-2h,0,0) - 4E(-h,0,0) + 6E(0,0,0) - 4E(+h,0,0) + E(+2h,0,0)}{h^4} \\
&\gamma_{yyyy} =\\
&-\frac{E(0,-2h,0) - 4E(0,-h,0) + 6E(0,0,0) - 4E(0,+h,0) + E(0,+2h,0)}{h^4} \\
&\gamma_{zzzz} =\\
&-\frac{E(0,0,-2h) - 4E(0,0,-h) + 6E(0,0,0) - 4E(0,0,+h) + E(0,0,+2h)}{h^4}
\end{align}

The mean (isotropic) second hyperpolarizability used in the $\beta/\gamma$ ratio is the average of the three diagonal components:
\begin{equation}
\bar{\gamma} = \frac{\gamma_{xxxx} + \gamma_{yyyy} + \gamma_{zzzz}}{3}
\label{eq:gamma_mean}
\end{equation}

\subsection{HOMO--LUMO Gap ($\Delta E$)}
\label{appendix:homolumo}

The HOMO--LUMO gap is computed from the canonical Hartree--Fock orbital energies $\{\epsilon_p\}$ obtained from the zero-field SCF calculation. For a closed-shell (restricted) molecule:
\begin{equation}
\Delta E = \epsilon_{\text{LUMO}} - \epsilon_{\text{HOMO}}
\end{equation}
where $\epsilon_{\text{HOMO}}$ is the energy of the highest occupied molecular orbital and $\epsilon_{\text{LUMO}}$ is the energy of the lowest unoccupied molecular orbital. For open-shell (unrestricted) calculations:
\begin{align}
\epsilon_{\text{HOMO}} &= \max\left(\epsilon_{\text{HOMO}}^\alpha,\; \epsilon_{\text{HOMO}}^\beta\right) \\
\epsilon_{\text{LUMO}} &= \min\left(\epsilon_{\text{LUMO}}^\alpha,\; \epsilon_{\text{LUMO}}^\beta\right)
\end{align}
where $\alpha$ and $\beta$ denote the two spin channels.

The gap is converted from atomic units (Hartree) to electron volts using $1\;\text{Ha} = 27.211\;\text{eV}$:
\begin{equation}
\Delta E \;(\text{eV}) = (\epsilon_{\text{LUMO}} - \epsilon_{\text{HOMO}}) \times 27.211
\label{eq:homo_lumo}
\end{equation}

By Koopmans' theorem~\cite{szabo:1996}, $-\epsilon_{\text{HOMO}}$ approximates the ionization potential and $-\epsilon_{\text{LUMO}}$ approximates the electron affinity within the Hartree--Fock framework.

\subsection{Total Energy per Atom}
\label{appendix:energy}

The total energy $E_{\text{total}}$ is the converged Hartree--Fock energy from the zero-field SCF calculation, which includes kinetic energy, nuclear--electron attraction, electron--electron repulsion (Coulomb and exchange), and nuclear--nuclear repulsion:
\begin{equation}
E_{\text{total}} = E_{\text{kinetic}} + E_{\text{ne}} + E_{\text{ee}} + E_{\text{nn}}
\end{equation}

The energy per atom objective normalizes by the total number of atoms $N_{\text{atoms}}$ (including implicit hydrogens):
\begin{equation}
\frac{E_{\text{total}}}{N_{\text{atoms}}}
\label{eq:energy_per_atom}
\end{equation}

This quantity is expressed in Hartree per atom. Physically reasonable molecules have negative values (e.g., $\approx -55$ to $-75$~Ha/atom for organic molecules in the 3-21G basis), while positive values indicate SCF convergence failures or severely strained geometries, and such molecules are excluded from the analysis.

\subsection{Computational Cost}
\label{appendix:cost}

Each molecule evaluation requires approximately 25--33 SCF calculations: 19 for the $\beta$ tensor (zero field, $\pm h$ on each axis, and all two-axis combinations), 6 additional for the $\gamma$ tensor ($\pm 2h$ on each axis), and the $\alpha$ calculation reuses the same zero-field and single-axis dipoles already computed for $\beta$. The HOMO--LUMO gap and total energy are extracted from the zero-field calculation at no additional cost. All SCF calculations use the same molecular geometry and basis set, differing only in the applied electric field.


\section{Evolutionary Computation Details}
\label{appendix:evo}

This appendix provides additional detail on the evolutionary computation and optimization concepts used in our experiments. The main text assumes familiarity with these methods; this section is intended for readers with backgrounds in physics, chemistry, or materials science.

\subsection{Evolutionary Algorithms}
\label{appendix:ea_general}

Evolutionary algorithms (EAs) are population-based optimization methods inspired by biological evolution~\cite{beyer:nc2002}. The general procedure is:

\begin{enumerate}
\item Initialization: Generate an initial population of $\mu$ candidate solutions (here, SMILES strings representing molecules).
\item Evaluation: Compute the fitness (objective value) of each individual using the quantum chemical calculations described in Appendix~\ref{appendix:calculations}.
\item Selection: Choose parents from the population based on their fitness. Higher-fitness individuals are more likely to be selected, creating selection pressure toward better solutions.
\item Variation: Apply genetic operators (mutation and/or crossover) to the selected parents to produce $\lambda$ offspring. In our experiments, we use mutation only (Section~\ref{appendix:mutations}), as crossover of SMILES strings rarely produces valid molecules.
\item Replacement: Form the next generation by selecting which individuals (from parents, offspring, or both) survive.
\item Termination: Repeat steps 2--5 for a fixed number of generations or evaluations.
\end{enumerate}

The key insight is that by iteratively selecting better solutions and introducing random variation, the population tends to improve over time, analogous to natural selection acting on heritable variation.

\subsection{Tournament Selection}
\label{appendix:tournament}

Tournament selection is used in our NSGA-II and $(\mu+\lambda)$ implementations. To select one parent, $k$ individuals are drawn uniformly at random from the population, and the best among them (by fitness or Pareto rank) is chosen. This process is repeated independently each time a parent is needed.

The tournament size $k$ controls selection pressure: larger $k$ increases the probability that the best individual in the population is selected, intensifying exploitation at the expense of exploration. We use $k = 3$.

Formally, given a population $P$ of size $|P|$, the probability that the best individual in $P$ is selected in a single tournament of size $k$ is:
\begin{equation}
\Pr(\text{best selected}) = 1 - \left(\frac{|P|-1}{|P|}\right)^k
\end{equation}
For $|P| = 40$ and $k = 3$, this is $\approx 7.4\%$, compared to $2.5\%$ with uniform random selection.

\subsection{$(\mu + \lambda)$ Elitist Selection}
\label{appendix:mu_lambda}

The $(\mu + \lambda)$ evolutionary strategy~\cite{beyer:nc2002} is a primitive single-objective evolutionary algorithm. At each generation:

\begin{enumerate}
\item The $\mu$ parents produce $\lambda$ offspring via mutation.
\item All $\mu + \lambda$ individuals (parents and offspring together) are ranked by fitness.
\item The top $\mu$ individuals survive to become parents of the next generation.
\end{enumerate}

The ``$+$'' notation indicates that parents compete with offspring for survival (elitist selection), meaning the best solution found so far is never lost. This contrasts with $(\mu, \lambda)$ selection, where parents are discarded and only offspring compete.

In our experiments, $\mu = \lambda = 20$, so 20 parents produce 20 offspring, and the best 20 of the combined 40 survive. Because this is a single-objective method, fitness is the $\beta/\gamma$ ratio. The simplicity of this scheme means it focuses search narrowly on high-$\beta/\gamma$ regions of chemical space, which explains both its excellent $\beta/\gamma$ scores and its poor performance on secondary objectives.

\subsection{NSGA-II}
\label{appendix:nsga2}

NSGA-II (Non-dominated Sorting Genetic Algorithm II) ~\cite{deb:tec2002} is a widely used multi-objective evolutionary algorithm. It maintains a population and uses two key mechanisms: non-dominated sorting and crowding distance, to simultaneously optimize multiple objectives while maintaining diversity along the Pareto front.

\subsubsection{Non-dominated Sorting}

The population is partitioned into successive \emph{fronts} $F_1, F_2, \ldots$ based on Pareto dominance (Equation~1 in the main text). The first front $F_1$ contains all non-dominated individuals: those not dominated by any other individual in the population. The second front $F_2$ contains individuals dominated only by members of $F_1$, and so on.

Formally, the front assignment proceeds as:
\begin{enumerate}
\item For each individual $\vec{x}_i$, count how many individuals dominate it ($n_i$) and record which individuals it dominates ($S_i$).
\item All individuals with $n_i = 0$ form $F_1$.
\item For each $\vec{x}_i \in F_1$, decrement $n_j$ for all $\vec{x}_j \in S_i$. Any $\vec{x}_j$ whose count reaches zero joins $F_2$.
\item Repeat until all individuals are assigned to fronts.
\end{enumerate}

During selection, individuals on front $F_1$ are always preferred over those on $F_2$, who are preferred over $F_3$, and so on. This ensures that the population converges toward the Pareto front.

\subsubsection{Crowding Distance}

When individuals share the same front, NSGA-II uses \emph{crowding distance} to prefer individuals in less crowded regions of objective space, promoting diversity along the Pareto front.

For a set of individuals on the same front, the crowding distance of individual $i$ is the sum of normalized distances between its two nearest neighbors in each objective dimension:
\begin{equation}
d_i = \sum_{m=1}^{M} \frac{f_m^{(i+1)} - f_m^{(i-1)}}{f_m^{\max} - f_m^{\min}}
\end{equation}
where $M$ is the number of objectives, $f_m^{(i+1)}$ and $f_m^{(i-1)}$ are the objective values of the neighbors of individual $i$ when sorted by objective $m$, and $f_m^{\max}$, $f_m^{\min}$ are the maximum and minimum values of objective $m$ in the front. Boundary individuals (those with the best or worst value in any objective) are assigned infinite crowding distance to ensure they are always preserved.

Individuals with larger crowding distance are preferred, as they represent less-explored regions of the trade-off surface. This prevents the population from collapsing to a small cluster on the Pareto front.

\subsubsection{NSGA-II Selection and Replacement}

The complete NSGA-II cycle is:
\begin{enumerate}
\item Generate $\lambda$ offspring from the current population of $\mu$ parents using binary tournament selection (Section~\ref{appendix:tournament}), where tournaments compare first by front rank (lower is better) and then by crowding distance (higher is better).
\item Combine parents and offspring into a pool of $\mu + \lambda$ individuals.
\item Apply non-dominated sorting to this combined pool.
\item Fill the next generation of size $\mu$ by adding complete fronts $F_1, F_2, \ldots$ in order. If a front does not fit entirely, its members are sorted by crowding distance and the most spread-out individuals are chosen.
\end{enumerate}

In our experiments, $\mu = \lambda = 20$, so the combined pool has 40 individuals and the top 20 survive each generation.

\subsection{MAP-Elites}
\label{appendix:mapelites}

MAP-Elites (Multi-dimensional Archive of Phenotypic Elites)~\cite{mouret:arxiv2015} is a single-objective quality diversity (QD) algorithm that maintains a structured archive of solutions rather than a traditional population. The archive is a grid indexed by \emph{measure functions} (also called behavior characteristics or descriptors) that quantify properties of interest distinct from fitness.

\subsubsection{Archive Structure}

The archive is a $k$-dimensional grid where each axis corresponds to a measure function $m_i: \mathcal{S} \to \mathbb{R}$. Each cell (bin) in the grid stores at most one solution: the highest-fitness solution whose measure values map to that cell. In our experiments, $k = 2$ with measures:
\begin{itemize}
\item $m_1$: number of heavy atoms (range $[5, 30]$)
\item $m_2$: number of bonds between heavy atoms (range $[4, 32]$)
\end{itemize}

The grid has $M \times M$ cells, with $M = 10$ (coarse) or $M = 20$ (fine). The bin index for measure $m_i$ is computed as in Equation~4 of the main text.

\subsubsection{Algorithm}

\begin{enumerate}
\item Initialization: Evaluate a set of random molecules and place each in the archive cell corresponding to its measure values. If a cell is already occupied, the molecule with higher fitness is retained.
\item Iteration: Repeat for a fixed number of evaluations:
\begin{enumerate}
\item Select a random occupied cell from the archive.
\item Apply mutation to the occupant to produce one offspring.
\item Evaluate the offspring's fitness and measure values.
\item Determine the offspring's target cell from its measures.
\item If the target cell is empty, place the offspring there.
\item If the target cell is occupied, the offspring replaces the occupant only if it has strictly higher fitness.
\end{enumerate}
\end{enumerate}

The key property of MAP-Elites is that it fills the archive with high-performing solutions across a diversity of measure values. Even if a new solution has low fitness, it will be preserved if it occupies a previously empty cell, encouraging exploration of the measure space. This is why MAP-Elites discovers structurally diverse molecules (varying atom and bond counts) even though it only optimizes $\beta/\gamma$.

\subsection{MOME}
\label{appendix:mome}

MOME (Multi-objective MAP-Elites)~\cite{pierrot:gecco2022} extends MAP-Elites to handle multiple objectives simultaneously by storing a local Pareto front in each archive cell rather than a single elite.

\subsubsection{Key Modification}

In standard MAP-Elites, each cell stores one solution and replacement is based on a scalar fitness comparison. In MOME, each cell stores a set of mutually non-dominated solutions (a Pareto front), and the addition rule becomes:

Given a new solution $\vec{x}$ mapping to cell $c$:
\begin{enumerate}
\item If cell $c$ is empty, add $\vec{x}$.
\item If cell $c$ is occupied, check dominance relationships between $\vec{x}$ and all current occupants $\{{\vec{x}_1, \ldots, \vec{x}_n}\}$:
\begin{itemize}
\item If any occupant dominates $\vec{x}$, discard $\vec{x}$.
\item Otherwise, add $\vec{x}$ and remove any occupants that $\vec{x}$ dominates.
\end{itemize}
\end{enumerate}

This means a cell can grow to contain many solutions, each representing a different trade-off among objectives. For example, one molecule in a cell might have high $\beta/\gamma$ but poor energy, while another has moderate $\beta/\gamma$ but excellent energy; both are retained because neither dominates the other.

\subsubsection{Parent Selection}

When generating offspring, MOME selects a random occupied cell and then a random solution from that cell's Pareto front. This ensures that diverse trade-off solutions across the archive contribute to producing new candidates.

\subsection{Simulated Annealing}
\label{appendix:sa}

Simulated annealing (SA)~\cite{dowsland:johnwiley1993, laarhoven:springer1987} is a non-evolutionary, single-solution metaheuristic inspired by the physical process of annealing in metallurgy, where a material is heated and then slowly cooled to reach a low-energy crystalline state.

\subsubsection{Algorithm}

SA maintains a single current solution $\vec{x}$ and iteratively generates a neighbor $\vec{x}'$ by applying a random perturbation. The acceptance rule is:

\begin{equation}
\Pr(\text{accept } \vec{x}') = \begin{cases} \quad
1 & \text{if } f(\vec{x}') \geq f(\vec{x}) \\ \quad\text{(improvement)} \\  \\ \quad
\exp\!\left(\dfrac{f(\vec{x}') - f(\vec{x})}{T}\right) & \quad \text{otherwise} \\ \quad \text{(Metropolis criterion)}
\end{cases}
\label{eq:metropolis}
\end{equation}
where $f$ is the fitness function (here, $\beta/\gamma$) and $T$ is the \emph{temperature} parameter. Better solutions are always accepted; worse solutions are accepted with a probability that depends on how much worse they are and the current temperature.

\subsubsection{Cooling Schedule}

The temperature decreases over time according to an exponential cooling schedule:
\begin{equation}
T_{t+1} = \alpha \cdot T_t
\end{equation}
where $\alpha \in (0, 1)$ is the cooling rate. We use $T_{\text{initial}} = 100$, $T_{\text{min}} = 0.01$, and $\alpha = 0.95$.

At high temperatures, most moves are accepted (favoring exploration of chemical space). As $T$ decreases, the acceptance probability for worse solutions drops, and the algorithm increasingly behaves as a greedy hill-climber (favoring exploitation of the current region).

\subsubsection{Neighbor Generation}

At each iteration, SA applies all seven mutation operators (Section~\ref{appendix:mutations}) to the current molecule, generating seven candidate neighbors. Each candidate is independently subjected to the acceptance criterion in Equation~\ref{eq:metropolis}, and the best accepted candidate becomes the new current solution. Over 285 iterations, this produces $285 \times 7 = 1{,}995$ evaluations.

\subsection{SMILES Mutation Operators}
\label{appendix:mutations}

Molecules are represented as SMILES (Simplified Molecular Input Line Entry System) strings~\cite{weininger:jcics1988}, which encode molecular graphs as linear text strings. Our mutation operators modify the underlying molecular graph rather than the string directly, ensuring chemical validity. Each operator:

\begin{enumerate}
\item Converts the SMILES string to an RDKit molecular graph object.
\item Applies the graph modification (adding/removing atoms, bonds, or rings).
\item Checks chemical validity (correct valences, connected graph, allowed elements).
\item Converts back to a canonical SMILES string.
\end{enumerate}

If a mutation produces an invalid molecule, it is retried up to 20 times with different random choices. If all retries fail, a different mutation operator is selected. Each offspring receives 1--3 sequentially applied mutations (except in SA, which applies one mutation per operator). The 7 mutation operators we employ are provided in Section~\ref{sec:mutations}.

\subsection{Hypervolume}
\label{appendix:hypervolume}

Hypervolume (HV)~\cite{zitzler:tec2003} is a widely used metric for evaluating the quality of a Pareto front in multiobjective optimization. It measures the volume of objective space that is simultaneously dominated by the Pareto front and bounded by a reference point.

\subsubsection{Definition}

Given a Pareto front $\mathcal{P} = \{\vec{f}(\vec{x}_1), \ldots, \vec{f}(\vec{x}_n)\}$ in $M$-dimensional objective space and a reference point $\vec{r}$ (a point that is dominated by all members of $\mathcal{P}$), the hypervolume is the Lebesgue measure (generalized volume) of the region dominated by $\mathcal{P}$ and bounded by $\vec{r}$:
\begin{equation}
\text{HV}(\mathcal{P}, \vec{r}) = \Lambda\!\left(\bigcup_{i=1}^{n} \{\vec{y} \in \mathbb{R}^M : \vec{f}(\vec{x}_i) \preceq \vec{y} \preceq \vec{r}\}\right)
\end{equation}
where $\Lambda$ denotes the Lebesgue measure and $\preceq$ denotes componentwise inequality (assuming minimization).

\subsubsection{Intuition}

In two objectives, the hypervolume is simply the area of the staircase-shaped region between the Pareto front and the reference point. In higher dimensions, it generalizes to a volume, hypervolume, etc. A larger hypervolume indicates a Pareto front that is both closer to the true optimum and more spread out across the trade-off surface.

The hypervolume has two desirable properties:
\begin{itemize}
\item Pareto compliance: If front $\mathcal{P}_1$ dominates front $\mathcal{P}_2$ (every point in $\mathcal{P}_2$ is dominated by some point in $\mathcal{P}_1$), then $\text{HV}(\mathcal{P}_1) \geq \text{HV}(\mathcal{P}_2)$.
\item Sensitivity to spread: A well-spread front covering more of the trade-off surface will generally have higher hypervolume than a clustered front.
\end{itemize}

\subsubsection{Normalization}

Because our four objectives have different scales and units ($\beta/\gamma$ is unitless and can reach thousands; $f_\alpha$ is in a.u.; $f_{\Delta E}$ is in eV; $E_{\text{tot}}/N_{\text{atoms}}$ is in Hartree), we normalize each objective to $[0, 1]$ before computing the hypervolume. The normalization ranges are given in Section~7.2 of the main text. After normalization, we use a reference point of $\vec{0}$ (the origin), which is dominated by all normalized solutions.

\subsection{Quality Diversity Score}
\label{appendix:qdscore}

The quality diversity score (QD score)~\cite{pugh:gecco2015} quantifies the overall quality of a QD archive for a single objective. It is the sum of fitness values across all occupied cells:
\begin{equation}
\text{QD}(A) = \sum_{c \in \text{occupied}(A)} f(x_c)
\label{eq:qd_score}
\end{equation}
where $A$ is the archive, $\text{occupied}(A)$ is the set of non-empty cells, and $f(x_c)$ is the fitness of the elite solution in cell $c$.

QD score increases when (a) more cells are filled (better coverage) and (b) the solutions in occupied cells have higher fitness (better quality). It thus captures both aspects of quality diversity in a single number. We compute QD scores separately for each objective, creating an archive for each where replacement is based on that specific objective.

\subsection{Multi-Objective Quality Diversity Score}
\label{appendix:moqdscore}

The multi-objective quality diversity score (MOQD score)~\cite{pierrot:gecco2022} extends QD score to the multi-objective setting. Because MOME cells contain Pareto fronts rather than single solutions, the quality of each cell is measured by its local hypervolume rather than a scalar fitness:
\begin{equation}
\text{MOQD}(A) = \sum_{c \in \text{occupied}(A)} \text{HV}(\mathcal{P}_c, \vec{r})
\label{eq:moqd_score}
\end{equation}
where $\mathcal{P}_c$ is the Pareto front in cell $c$ and $\vec{r}$ is the reference point used for hypervolume calculation.

MOQD score captures three desirable properties simultaneously:
\begin{itemize}
\item Coverage: More occupied cells contribute more terms to the sum.
\item Quality: Better Pareto fronts within cells have higher local hypervolumes.
\item Trade-off diversity: Cells with well-spread Pareto fronts (representing diverse trade-offs) have higher hypervolumes than cells with clustered fronts.
\end{itemize}

Although MOQD score was introduced for MOME, it can be applied to any algorithm by post-hoc construction of a MOME-style archive from the stream of all generated molecules, which is what we do for the non-MOME algorithms in our experiments.

\subsection{Global Hypervolume vs.\ MOQD Score}
\label{appendix:hv_vs_moqd}

These two metrics capture different aspects of performance and can rank algorithms differently:

\begin{itemize}
\item Global hypervolume computes a single Pareto front across \emph{all} molecules generated by an algorithm and measures its hypervolume. It rewards finding molecules with good trade-offs regardless of their structural diversity. An algorithm can achieve high global hypervolume by finding a small cluster of excellent molecules in one region of chemical space.
\item MOQD score sums local hypervolumes across the \emph{measure space} archive. It rewards finding good trade-offs in \emph{many different structural niches}. An algorithm must produce diverse molecules (varying atom/bond counts) with good objective trade-offs to score highly.
\end{itemize}

This distinction explains why $(\mu + \lambda)$ achieves moderate global hypervolume (driven by extremely high $\beta/\gamma$) but poor MOQD score (all its good molecules cluster in a small region of measure space), while MOME$_F$ excels at both.

\subsection{Pareto Dominance}
\label{appendix:pareto_example}

To illustrate Pareto dominance concretely, consider three molecules evaluated on two objectives: maximize $\beta/\gamma$ and minimize $E_{\text{total}}/N_{\text{atoms}}$.

\begin{center}
\begin{tabular}{lcc}
\toprule
Molecule & $\beta/\gamma$ (maximize) & $E_{\text{total}}/N_{\text{atoms}}$ (minimize) \\
\midrule
A & 500 & $-60$ \\
B & 300 & $-70$ \\
C & 200 & $-55$ \\
\bottomrule
\end{tabular}
\end{center}

\begin{itemize}
\item A vs.\ B: A has higher $\beta/\gamma$ (better) but higher $E_{\text{total}}/N_{\text{atoms}}$ (worse). Neither dominates the other---they represent different trade-offs.
\item A vs.\ C: A has higher $\beta/\gamma$ (better) and lower $E_{\text{total}}/N_{\text{atoms}}$ (better). A dominates C ($A \prec C$).
\item B vs.\ C: B has higher $\beta/\gamma$ (better) and lower $E_{\text{total}}/N_{\text{atoms}}$ (better). B dominates C ($B \prec C$).
\end{itemize}

The Pareto front is $\{A, B\}$: these molecules represent the best known trade-offs between the two objectives, and neither can be improved in one objective without worsening the other. Molecule C is dominated and would be discarded by NSGA-II in favor of A or B.

With four objectives (as in our experiments), the dominance comparison extends componentwise across all four dimensions, and the Pareto front becomes a surface in four-dimensional objective space.


\section{Statistical Methods}
\label{appendix:statistics}

This appendix describes the statistical methods used to analyze and compare algorithm performance across 20 independent runs with different random seeds.

\subsection{Use of Median Instead of Mean}
\label{appendix:median}

We report \emph{median} scores rather than means for two reasons:

\begin{enumerate}
\item Non-normal distributions: The distributions of objective scores and hypervolume values across runs are typically right-skewed with heavy tails. Some runs discover exceptionally good molecules (high outliers), while most runs cluster at more moderate values. The mean is sensitive to these outliers and can misrepresent typical algorithm performance.

\item Rigidity: The median is the 50th percentile, exactly half of the runs perform better and half worse. It is a robust measure of central tendency that is not inflated by a single exceptional run. For example, if 19 of 20 $(\mu+\lambda)$ runs produce moderate $\beta/\gamma$ ratios, but one run finds an anomalously high value, the mean would be substantially inflated while the median would not.
\end{enumerate}

\subsection{Box-and-Whisker Plots}
\label{appendix:boxplots}

Figure~2b in the main text uses box-and-whisker plots to display the distribution of final hypervolume scores across all 20 runs. The components are:

\begin{itemize}
\item Box: Spans from the first quartile ($Q_1$, 25th percentile) to the third quartile ($Q_3$, 75th percentile). The height of the box is the interquartile range $\text{IQR} = Q_3 - Q_1$, which contains the middle 50\% of the data.
\item Center line: The median ($Q_2$, 50th percentile).
\item Whiskers: Extend to the most extreme data points within $1.5 \times \text{IQR}$ of the nearest quartile. Specifically, the lower whisker extends to $\min(x_i : x_i \geq Q_1 - 1.5 \cdot \text{IQR})$ and the upper whisker to $\max(x_i : x_i \leq Q_3 + 1.5 \cdot \text{IQR})$.
\item Outliers: Points beyond the whiskers are plotted individually. These represent runs with unusually high or low performance.
\end{itemize}

\subsection{Number of Runs and Reproducibility}
\label{appendix:nruns}

We use 20 independent random seeds per algorithm. This number was chosen as a balance between statistical power and computational cost (each run evaluates ${\sim}2{,}000$ molecules, each requiring ${\sim}30$ SCF calculations). With 20 runs, the median is the average of the 10th and 11th ranked values, providing a stable estimate of typical performance. The quartiles and box plots give a sense of the variability across runs.

Because the stochastic elements differ across algorithms (random initialization, random parent selection, random mutation choices, and SA's probabilistic acceptance), the use of multiple independent runs is essential to distinguish genuine algorithmic differences from run-to-run noise.

\subsection{Evaluation Budget and Fair Comparison}
\label{appendix:budget}

All algorithms are compared at approximately equal evaluation budgets (${\sim}2{,}000$ molecules) rather than equal wall-clock time or equal generations. This is the standard practice in evolutionary computation because the fitness evaluation (here, quantum chemical calculation) dominates the computational cost. The algorithmic overhead of selection, sorting, and archive management is negligible by comparison. One molecule evaluation takes many minutes of CPU time for the HF/3-21G finite-field procedure (scaling to the number of atoms in the molecule), while most of the EA operations take microseconds. The specific budgets for each algorithm are in the main text Section \ref{sec:settings}.

\end{document}